\documentclass{article}
\usepackage{float}

\usepackage{multirow}
\usepackage{longtable}

\usepackage{PRIMEarxiv}

\usepackage[utf8]{inputenc} 
\usepackage[T1]{fontenc}    
\usepackage{hyperref}       
\usepackage{url}            
\usepackage{booktabs}       
\usepackage{amsfonts}       
\usepackage{nicefrac}       
\usepackage{microtype}      
\usepackage{lipsum}
\usepackage{fancyhdr}       
\usepackage{graphicx}       
\usepackage{subcaption}
\usepackage{bm}
\usepackage{soul}
\usepackage{comment}

\usepackage{tcolorbox}

\tcbuselibrary{listings} 
\newtcolorbox[auto counter]{myalgorithm}[2][]{
  floatplacement=htbp,
  float,
  title=Algorithm~\thetcbcounter: #2,
  label={#1},
  enhanced
}

\graphicspath{{media/}}     

\DefineNamedColor{named}{Purple}{cmyk}{0.45,0.86,0,0}

\DefineNamedColor{named}{JungleGreen} {cmyk}{0.99,0,0.52,0}

\DefineNamedColor{named}{orange} {rgb}{1,0.55,0}

\usepackage{amssymb}
\usepackage{lipsum}
\usepackage{amsmath}
\DeclareMathOperator*{\argmin}{argmin}
\usepackage{natbib}
\usepackage{url}

\usepackage{breakurl}

\title{Integration of Local and Global Surrogates for Failure Probability Estimation
}

\author{
  Audrey Gaymann\\
  University of Colorado, Boulder \\
  \texttt{audrey.gaymann11@colorado.edu} \\
  \And
  Juan M. Cardenas\\
  University of Colorado, Boulder \\
  \texttt{Juan.CardenasCardenas@colorado.edu} \\
  \And
  Sung Min Jo\\
  University of Central Florida\\
  \texttt{sungmin.jo@ucf.edu}\\
  \And
  Marco Panesi\\
  University of California, Irvine\\
  \texttt{mpanesi@uci.edu}\\
  \And
  Alireza Doostan\\
  University of Colorado, Boulder\\
  \texttt{doostan@colorado.edu}\\
   }

\begin{document}
\maketitle

\begin{abstract}
This paper presents the development of an algorithm, termed the Global-Local Hybrid Surrogate (GLHS), designed to efficiently compute the probability of rare failure events in complex systems. The primary goal is to enhance the accuracy of reliability analysis while minimizing computational cost, particularly for high-dimensional problems where traditional methods, such as Monte Carlo simulations, become prohibitively expensive. The proposed GLHS builds upon the foundational work of \cite{LI20108966,LI20118683} by integrating an adaptive strategy based on the General Domain Adaptive Strategy \citep{doi:10.1137/22M1472693}. The algorithm aims to approximate the failure domain of a given system, defined as the region in the input domain where the system transitions from safe to failure modes, described by a limit state surface. This failure domain is not explicitly known and must be learned iteratively during the analysis. The method employs a buffer zone, defined as the region surrounding the limit state surface. Within this buffer zone, Christoffel Adaptive Sampling is utilized to select new samples for constructing localized surrogate models, which are designed to refine the approximation in regions critical to failure probability estimation.
The iterative process proceeds until convergence is reached. This results in a hybrid methodology that integrates a global surrogate to capture the overall trend with local surrogates that concentrate on critical regions near the limit state function. By adopting this strategy, the GLHS method balances computational efficiency with accuracy in estimating the failure probability.

\end{abstract}


\section{Introduction}
\label{introduction}

Reliability analysis plays a critical role in evaluating the performance and safety of complex engineering systems, particularly under uncertain conditions. These uncertainties can arise from aleatory sources, such as inherent variability in physical processes, or epistemic sources, such as limited knowledge of system parameters. Addressing these challenges is particularly important in high-dimensional problems, where computational resources are often a limiting factor. Hybrid reliability analysis (HRA), which integrates the effects of both aleatory and epistemic uncertainties, has emerged as a promising approach to balance precision and computational efficiency, especially in recent years for structural analysis applications \cite{WANG20171451,https://doi.org/10.1111/j.1539-6924.2009.01221.x, GAO2011643, CHOWDHURY2009753}. Recent advances have focused on developing surrogate modeling techniques and adaptive sampling strategies to reduce the reliance on expensive, high-fidelity simulations, while maintaining accurate failure probability estimations. 

Let \(g_T(\bm{x})\) denote the limit state function for input variables \(\bm{x} \in \mathbb{R}^d\) following distribution $\tau$.  Failure is defined as the event in which the limit state function becomes non-positive, i.e., \(g_T(\bm{x}) \le 0\). The corresponding probability of failure is therefore given by
\[
P_f = \mathbb{P}\!\left(g_T(\bm{x}) \le 0\right).
\]

A variety of methods have been proposed in the literature to estimate $P_f$, with Monte Carlo simulation (MCS) being one.

MCS \cite{MCmethods,MCsimulation} is widely regarded for its ability to deliver accurate results across diverse problems and dimensions without relying on specific assumptions about the underlying system. However, its computational cost becomes prohibitive when estimating small probabilities, as the required sample size scales approximately as $\mathcal{O}(\frac{1}{P_f})$ \cite{AU2001263}. Given the high cost of generating samples, it is crucial to minimize their use, which highlights the importance of surrogate modeling as an alternative for estimating failure probabilities.

Surrogate modeling methodologies offer promising solutions by employing various approximation techniques, such as polynomial, radial basis function, and Gaussian process models, to reduce the computational burden in reliability analysis \cite{RAJASHEKHAR1993205}. These methods focus on approximating the relationship between input variables and system responses, thereby enabling efficient failure probability estimation without requiring exhaustive simulations. Among these approaches, Polynomial Chaos Expansions (PCE), \cite{Ghanem1991,xiu2002wiener}, have been widely utilized for reliability analysis (\cite{PAFFRATH2007263}). However, high-dimensional uncertain inputs pose challenges for PCEs due to the curse of dimensionality issue, where an accurate surrogate model requires a significantly large number of model evaluations, \cite{doostan2009least,doostan2011non}.  

Alternative methodologies such as the First Order Reliability Method (FORM) and Second Order Reliability Method (SORM) are prevalent in the literature, especially in the structural reliability applications \cite{RACKWITZ2001365, SCHUELLER1987293}. FORM seeks to estimate the failure probability of a system via a linear approximation of $g_T(\bm{x} = 0)$, i.e., first-order Taylor series expansion around the most probable point in the space, often referred to as the design point or most probable point. The failure probability is subsequently determined by calculating the standard normal distribution function, which measures how far the point is from the origin in the standard normal space. This distance is referred to as the reliability index or safety margin, representing the likelihood of failure. A smaller distance indicates a higher probability of failure, while a larger distance suggests greater safety and lower failure risk. Second Order Reliability Method (SORM) \cite{SCHUELLER2004463,app11010346} is an extension of FORM that provides improved accuracy by accounting for the curvature of the limit state function. Unlike FORM, which relies on a linear approximation, SORM uses a quadratic approximation (second-order Taylor series expansion) to characterize the limit state function near the design point. By considering the curvature of the limit state surface, i.e., $\{\bm x: g_T(\bm x)=0\}$, SORM offers more accurate estimates of the failure probability, especially for limit state functions exhibiting significant nonlinearity. SORM typically involves a higher computational cost than FORM due to the need for additional calculations --- related to the Hessian matrix at the design point --- to account for second-order effects.

The objective of recent studies --- as well as this work --- is to enhance the accuracy of a surrogate model in estimating the failure probability, $\mathbb{P}(g_S(\bm{x})\leq 0) \approx P_f$, where \(g_{S}(\bm{x})\) denotes a (static) global surrogate of the function $g_T$. Our work is inspired by the studies of \cite{LI20108966, LI20118683}, which integrate surrogate modeling with Monte Carlo sampling within a region encompassing the limit state surface, $\{\bm{x}:|g_T (\bm{x})| \leq \eta \}$ for a small threshold $\eta$. This region shall be referenced as the buffer zone.
The Global-Local Hybrid Surrogate (GLHS) method proposed in this paper constructs local surrogates in addition to a global surrogate to more accurately approximate the limit state function near or within the buffer zones. This approach allows for a more cost-effective estimation of the probability of failure by replacing traditional Monte Carlo sampling with a targeted, reduced number of samples. While the global surrogate captures the overall trend of the entire domain, it is less effective at predicting local behavior at or near the limit state surface. Therefore, we aim to enhance its accuracy by utilizing local surrogates, which are better suited for estimating trends in specific regions. These local surrogates are constructed using a small set of samples, to achieve $\mathbb{ \ P(}g_{S}(\bm{x}) \leq 0) \approx \mathbb{\ \ P(}g_{T}(\bm{x}) \leq 0) $. 

Building an effective local surrogate presents two primary challenges: accurately modeling the limit state surface and efficiently selecting samples to minimize the required computational cost. To overcome these challenges, the buffer zone is introduced to isolate regions near the limit state surface that demand higher accuracy. Within this zone, the General Domain Adaptive Strategy \citep{doi:10.1137/22M1472693} employs Christoffel Adaptive Sampling (CS) to optimally select sample points, ensuring the local surrogate effectively approximates the target function while reducing the reliance on model evaluations.

The rest of the manuscript is organized as follows. 
Section~\ref{sec:background} reviews existing approaches to failure probability 
estimation, including the non-iterative and iterative methods of 
\cite{LI20108966, LI20118683}, as well as background on CS Sampling. 
Section~\ref{section:LGM} presents the proposed GLHS methodology in detail. 
Numerical examples are provided in Section~\ref{sec:ne}, and 
conclusions are given in Section~\ref{sec:con}.

\section{Background}
\label{sec:background}

This section provides an overview of existing methodologies for failure probability estimation, focusing on hybrid approaches that combine different techniques to balance computational efficiency and accuracy. Specifically, we discuss the non-iterative and iterative methods developed by \cite{LI20108966, LI20118683}, which motivated the present study. These methods aim to enhance the accuracy of failure probability estimation compared to standalone surrogate models while reducing the computational cost of model evaluations inherent in traditional MCS schemes.

A hybrid approach in this context refers to combining global and local modeling techniques to achieve a more accurate representation of the failure domain. The methods proposed by \cite{LI20108966, LI20118683} integrate surrogate models with Monte Carlo sampling, focusing on refining the surrogate model in the proximity of the limit state surface. However, their hybrid methodology differs from the GLHS method presented in this study. While Li et al.'s methods use pre-defined strategies for refining the surrogate model, the GLHS method employs a domain-adaptive approach with CS to iteratively refine local surrogates within dynamically identified buffer zones. 

In the following subsections, we summarize the key elements of the non-iterative and iterative methods by Li et al., along with an introduction to CS (CS) methodology, which forms a critical component of the GLHS approach. For clarity, a detailed nomenclature defining the terms and symbols used throughout this paper is provided in Appendix \ref{appendix:nomenclature}.

\subsection{Monte Carlo simulation}
\label{subsec:MC}

The most direct approach for estimating \(P_f\) involves generating a sufficiently large sample set \(\mathcal{S}=\{\bm{x}_i\}_{i=1}^{m_c}\), with each realization drawn from the prior distribution \(\bm{x}_i \sim \tau\), where \(\tau\) characterizes the probabilistic model of the inputs \(\bm{x} \in \mathbb{R}^d\).
Subsequently, the model is evaluated at these samples to derive the Monte Carlo estimate of the failure probability 
\begin{equation}
\label{eq:PFMC}
    {P_f^{MC} = \frac{1}{m_{c}} \sum_{i = 1}^{m_{c}}{\mathbb{I}_{g_T\leq0}(\bm{x}_i)}}.
\end{equation}
Here, $\mathbb{I}_{g_T\leq0}$ represents the indicator function, taking the value 1 when $g_T(\bm{x})\leq 0$ and 0 otherwise. Monte Carlo methods are widely regarded for their generality and asymptotic accuracy across various problem settings. However, they suffer from slow convergence, with an error rate of $\mathcal{O}(\frac{1}{\sqrt{m_c}})$. As a result, a large number of samples is required to obtain an accurate estimate, particularly when the probability of failure is small. In such cases, the number of required samples scales as $\mathcal{O}(\frac{1}{P_f})$, making standard Monte Carlo approaches computationally impractical for rare event estimation. This motivates the need for surrogate modeling and variance reduction techniques to reduce the number of model evaluations.

\subsection{Surrogate modeling}
\label{subsec:RS}
This approach revolves around constructing a surrogate $g_S$ by fitting it to $m_0$ samples of $g_T$, i.e., the samples $\{\bm{x}_i,g_T(\bm{x}_i)\}_{i=1}^{m_0}$. The objective of the surrogate method is to approximate the limit state function,
\begin{equation}
    {g_S(\bm{x}) \approx g_T(\bm{x})}.
\end{equation}
As compared to the original model, sampling a surrogate is significantly less expensive since no additional simulations are needed beyond those used to build $g_S(\bm{x})$. The probability of failure is then estimated using
\begin{equation}
    {P_f^{RS} = \frac{1}{m_{c}} \sum_{i = 1}^{m_{c}}{\mathbb{I}_{g_S\leq0}(\bm{x}_i)}}.
\end{equation}
While the surrogate model offers a computationally efficient alternative to directly evaluating the limit state function, there are inherent challenges associated with its use. The accuracy of the surrogate model heavily depends on the quality and number of initial samples $m_0$ used to fit the surrogate $g_S$. If the sampling is insufficient or poorly distributed, the surrogate may not capture the true behavior of the original model, leading to biased or unreliable failure probability estimates. Additionally, surrogate models are particularly sensitive to the complexity and nonlinearity of the underlying model. Furthermore, the approach assumes that once the surrogate model is constructed, no additional simulations are needed. However, in practice, the surrogate might require recalibration or refinement, particularly in areas where the initial samples do not provide sufficient coverage. Thus, while surrogate modeling can significantly reduce computational costs, it is crucial to carefully assess its limitations and ensure that the model's accuracy remains high, especially in the context of small failure probability estimation. Throughout our study, this method serves as an initial step, acting as the baseline against which we seek improvement.  

In this work, both the global surrogate and the subsequent local surrogates are constructed using PCE. However, it is important to note that the global surrogate model could be replaced with an alternative surrogate modeling technique, depending on the application or data characteristics.

A PCE approximates the output quantity of interest (QoI) as a spectral expansion of orthogonal polynomials in terms of the input random variables, and takes the general form:
\begin{equation}
    g_S(\bm{x}) \approx \sum_{i=1}^{N} c_i \, \psi_i(\bm{x}),
\end{equation}
where \( \psi_i(\bm{x}) \) are multivariate orthogonal polynomial basis functions, $N$ the total number of basis functions and \( c_i \) are the corresponding coefficients.

The construction of a PCE requires knowledge of the prior distribution of the input variables $\tau$. This prior determines the appropriate family of orthogonal polynomials (e.g., Hermite for Gaussian inputs, Legendre for uniform inputs). Once the basis is chosen, the coefficients \( c_i \) are computed either via projection (intrusive methods) or regression techniques (non-intrusive methods), depending on the available data and computational resources. 

\subsection{Hybrid surrogate and full model approaches}
\label{subsec:NIM}

\paragraph{Non-iterative method:} Proposed by \cite{LI20108966}, this method begins by constructing a surrogate aimed at approximating $g_T$. A threshold $\eta$ is identified or prescribed, representing a small positive value. This threshold determines a buffer zone around the limit state surface, $\mathcal{S}_\eta = \{\bm{x}:|g_S(\bm{x})| \leq \eta\}$. Sampling of the system is then conducted based on the probability distributions of each input. A sampling over the entire domain, which is not a Monte Carlo sampling strategy, will have a total number of samples referenced as $m$, with $m \ll m_c$, yielding the samples $\{\bm{x}_i\}_{i=1}^{m}$, $\bm{x}_i\in\mathbb{R}^d$. The limit state function is only computed for values falling within the buffer zone, leading to the definition of the probability of failure as:
\begin{equation}
    P_f^{NI} = \frac{1}{m} \sum_{i=1}^{m}\mathbb{I}_{\mathcal{S}_f}(\bm{x}_i).
\end{equation}
Here, the failure domain $\mathcal{S}_f$ is characterized by:
\begin{equation}
    \mathcal{S}_f = \{\bm{x}: g_S(\bm{x}) < - \eta \} \cup \left\{ \{\bm{x}:|g_S(\bm{x})| \leq \eta \} \cap \{\bm{x}:g_T(\bm{x}) \leq 0 \} \right\} .
\end{equation}

One notable drawback of this method lies in its reliance on the chosen value of $\eta$ as a function of the $\ell^p$-norm of the error of the surrogate model, for some $p \geq 1$, \cite{LI20108966}. While computing the original model solely within the buffer zone is less costly than a Monte Carlo solution, setting the buffer zone too conservatively increases the method's computational cost due to a higher number of samples required. Sampling the buffer zone is essential in this methodology, as all samples within the buffer zone require model evaluations. Hence, the method relies on a surrogate that fits the data well in a strong norm, such as the $L^\infty$ norm, to ensure a small buffer zone. 

\paragraph{Iterative method:}
To reduce the sensitivity to the parameter \(\eta\), an alternative procedure for estimating the
failure probability was proposed in \cite{LI20118683}. The approach begins by sampling input variables
from their prior distribution \(\tau\), after which a surrogate model is constructed. The samples are then
sorted in ascending order of \(|g_S(\bm{x})|\), and subsequently divided into groups of size
\(\delta m \in \mathbb{N}\). Let \(K\) denote the total number of groups. Iteratively, for each group
\(k = 1,\dots, K\), $g_T$ is evaluated for the samples indexed from
\((k-1)\,\delta m + 1\) to \(k\,\delta m\). Using these evaluations, the estimate of the failure
probability is updated, and the process continues until convergence is achieved.

One potential concern with this algorithm is that, under certain convergence behaviors, it may begin to resemble a Monte Carlo approach if too many samples end up requiring model simulations. As before, the effectiveness of the method depends on the quality of the surrogate, which must be accurate enough to limit the number of required $g_T$ evaluations.

\subsection{Christoffel Adaptive Sampling (CS)}
\label{subsec:CS}

The techniques discussed in Sections~\ref{subsec:NIM} and~\ref{subsec:NIM} highlight the challenge of balancing accuracy and computational cost in failure probability estimation. Both the non-iterative and iterative approaches discussed rely on Monte Carlo sampling of the buffer region, i.e., following $\tau$ conditioned on the buffer zone, which can become expensive when a large number of model simulations are required, for instance, when the size of the buffer zone is large. 

Alternatively, this study relies on local surrogates to avoid direct Monte Carlo sampling of the buffer zone. To that end, it is essential to efficiently sample inputs to minimize the number of model evaluations. This section introduces CS as an instance of such a strategy specifically for the construction of GLHS models.

CS selects optimal samples based on the so-called Christoffel function associated with the finite-dimensional approximation space used to construct the surrogate. This function quantifies how well a point is represented by a given polynomial basis, inherently favoring regions where the associated regression problem lends itself to improved convergence, as compared to sampling from the original distribution of inputs. For more details about such convergence properties, the interested reader is referred to \cite{Hampton_2015, HAMPTON2015363,HADIGOL2018382,adcock2025optimalsamplingleastsquaresapproximation, Narayan_2016, cohen2016optimalweightedleastsquaresmethods}. CS is especially advantageous when building local PCE surrogates within GLHS: while the prior distribution $\tau$ of inputs is known, the distribution of inputs over the buffer zone -- where the local surrogate is built -- may be different. 

Consider a grid $\mathcal{X}_l$, a finite set of samples drawn over the support of the local surrogate -- a subset of the continuous input domain $\mathcal{D}$ -- used to evaluate the QoI. 
The subscript $l$ is introduced here for clarity, as GLHS is iterative and subsequent iterations will use incremented subscripts to denote updated grids and domains of the local surrogates. To define the sampling measure over the support of each local surrogate, we assume that we can construct a new sampling measure based on $\tau$. We define the measure $\nu$ as 
\begin{equation}
    \text{d}\nu (\bm{x}) = w(\bm{x}) \text{d}\tau_l(\bm{x}), \quad\bm{x}\in\mathcal{X}_l,
\end{equation}
where $w(\bm{x})$ is a weight function adjusting the original probability distribution $\tau$ to prioritize regions of interest within the discrete domain $\mathcal{X}_l$ according to the Christoffel function.  
The measure $\tau_l$ is the restriction of $\tau$ to the domain $\mathcal{X}_l$, defined as 
\begin{equation}
\text{d}\tau_{l}(\bm{x}) 
= \frac{\mathbb{I}_{\mathcal{X}_l}(\bm{x})}{\int_{\mathcal{X}_l}^{}{\text{d}\tau(\bm{x})}}\text{d}\tau(\bm{x}),
\quad \bm{x} \in \mathcal{D}=[-1,1]^d,    
\end{equation}
where \(\mathbb{I}_{\mathcal{X}_l}(\bm{x})\) is the indicator function of $\bm{x}$ in $\mathcal{X}_l$. This is a nonempty set by construction, and $\tau$ is supported on $\mathcal{X}_l$. Therefore, $\tau_{l}$ is well defined. Next, we consider a finite-dimensional approximation space defined as 
\begin{equation*}
    P_{\phi} = \mathrm{span}\{\phi_1,\ldots,\phi_N\}\subset L^2(\mathcal{X}_l,\tau_l),
\end{equation*}
where $\{\phi_1,\ldots,\phi_N\}$ is an orthonormal basis with respect to $\tau_l$ and $N$ is the number of basis functions in the approximation space $P_\phi$. Now, we can define the reciprocal of the Christoffel function of the space $P_{\phi}$ as 
\begin{equation}
K_C(P_\phi)(\bm{x}) 
= \frac{1}{N}\sum_{i = 1}^{N}\left| \phi_{i}(\bm{x}) \right|^{2},
\end{equation}
also known as the sum of the diagonal of the kernel generated by $P_\phi$. Note that the Christoffel function is defined as $1/\sum_{i=1}^N|\phi_i(\bm{x})|^2$.  
Lower values of \(K_C(P_\phi)(\bm{x})\) indicate regions where more samples are needed. Therefore, we define the weights in our sampling measure as 
\begin{equation}
\label{eqn:weights}
    w(\bm{x})=K_C(P_\phi)(\bm{x}),
\end{equation}
leading to the sampling measure 
\begin{equation}
\text{d}\nu(\bm{x}) 
= K_C(P_\phi)(\bm{x})\text{d}\tau_{l}(\bm{x}) 
= \frac{1}{N}\sum_{i = 1}^{N}\left| \phi_i(\bm{x}) \right|^{2}\text{d}\tau_l(\bm{x})
,\ \bm{x} \in \mathcal{D}.
\end{equation}
While Christoffel functions and optimal sampling strategies are classically defined for a continuous probability measure on the domain $\mathcal{D} \subset \mathbb{R}^d$, the GLHS algorithm operates entirely on a finite discrete representation of the domain. Indeed, all evaluations of the surrogate and all identification of buffer zones are performed on the grid
\begin{equation}
    \mathcal{X}_l = \{\tilde{\bm x}_i\}_{i=1}^{m_d} \subset \mathcal{D},
\end{equation}
which is a discrete representation of a local subdomain of  $\mathcal{D}$ using $m_d\gg1$ samples. Consequently, the continuous restricted measure $\tau|_{S_\eta}$ is replaced by a discrete empirical measure supported on $\mathcal{X}_l$. This discretization is essential for GLHS, because the buffer zone is only available through evaluations of $g_S$ on a finite grid. Christoffel sampling also requires inner products and orthonormal systems defined with respect to a finite measure. Thus, the use of the discrete measure is not merely an approximation but a necessary ingredient for implementing the Christoffel sampling strategy within the iterative algorithm.

The sampling measure is well used in the context of regression problems with an orthonormal system. However, the domain represented by $\mathcal{X}_l$ is in general non-Cartesian. Instead, we construct an orthonormal basis on the domain represented by $\mathcal{X}_l$. Let us start with a non-orthonormal basis $\{\psi_1,\ldots,\psi_N\}$ corresponding to $\mathcal{X}_l$, consisting of multi-variate Legendre polynomials (with hyperbolic cross truncation \cite{hypcross}) for the case of uniform $\tau$. Using $\{\psi_1,\ldots,\psi_N\}$, we seek to construct an orthonormal basis over the domain represented by $\mathcal{X}_l$. To this end, we define the corresponding discrete sampling measure as 
\begin{equation}
    \text{d}\tau_{l}(\bm{x}) \approx \sum_{i = 1}^{m_d}\rho_{i}\delta\left(\bm{x} - \tilde{\bm{x}}_{i} \right),
\end{equation}
where $\tilde{\bm{x}}_i$ is the $i$-th point in the grid $\mathcal{X}_l$, $\rho_{i} = 1/m_d$, and $\delta$ is the Dirac function.
 
Although the Legendre polynomials $\{\psi_1,\ldots,\psi_N\}$ are orthonormal on the continuous domain $\mathcal{D}$ -- when $\tau$ is uniform -- they are not orthonormal on the discrete grid $\mathcal{X}_l$ used in GLHS. To obtain an orthonormal basis on $\mathcal{X}_l$, following \cite{adcock2019nearoptimalsamplingstrategiesmultivariate,doi:10.1137/22M1472693}, we construct the measurement matrix $\bm{B}$ given by 
\begin{equation}
\label{eqn:B_mat}
\bm{B} 
= \left\lbrace \frac{1}{\sqrt{m_d}}\psi_{j}(\tilde{\bm{x}}_{i}) \right\rbrace_{i,j = 1}^{m_d,N} \in \mathbb{R}^{{m_d} \times N}.
\end{equation}
Note that, if $\bm{B}$ does not have full rank ($\text{rank}(\bm{B})<N$), $m_d$ should be increased by drawing additional samples of $\bm{x}$. Then we compute the QR decomposition of $\bm{B}=\bm{Q}\bm{R}$, where $\bm{Q} \in \mathbb{R}^{m_d \times N}$ and $\bm{R} \in \mathbb{R}^{N \times N}$. The basis $\{\phi_1,\ldots,\phi_N\}$ defined as
\begin{equation}
\phi_{i}(\bm{x}) 
= \sum_{j = 1}^{i}\left( \bm{R}^{- \top} \right)_{ij}\psi_{j}(\bm{x})
,\quad i = 1,\ldots,N,
\end{equation}
is an orthonormal for $P_\phi$ in $L^{2}\left( \mathcal{X}_l,\tau_{l} \right)$. 

Finally, the Christoffel sampling measure $\nu$ is expressed as
\begin{equation}
\text{d}\nu(\bm{x}) 
= \sum_{k = 1}^{m_d}{\sum_{j = 1}^{N}{\left| q_{kj} \right|^{2}\delta\left(\bm{x} - \tilde{\bm{x}}_{i} \right)\text{d}\bm{x}}}.
\end{equation}
Hence, $\nu$ is the  discrete probability measure with $\bm{x} \sim \nu$. Sampling from
$\nu$ is equivalent to drawing an integer $t$ randomly from $\text{\{}1,\ldots,m_d\text{\}}$ based on distribution $\text{\{}\sum_{j = 1}^{N}\left| q_{kj} \right|^{2}\text{\}}_{k = 1}^{m_d}$ and then setting $\bm{x} = \tilde{\bm{x}}_{t}$.

For further discussions related to the stability of the QR approach and its improvement for constructing the orthonormal basis $\{\phi_1,\ldots,\phi_N\}$, we refer the reader to \cite{adcock2019nearoptimalsamplingstrategiesmultivariate}. 

\section{Global-Local Hybrid Surrogate (GLHS) Method}
\label{section:LGM}

In this section, we introduce the proposed GLHS framework for estimating probabilities of failure while minimizing the number of $g_T$ evaluations. Building upon the background presented in Section~\ref{sec:background}, the GLHS method combines a global surrogate model, which captures the overall behavior of the limit state function over the full input domain, with a sequence of local surrogate models that are adaptively constructed in regions most critical to failure probability estimation.

The central motivation of GLHS is that global surrogate models alone may lack sufficient accuracy near the limit state surface, where even small local approximation errors can lead to significant bias in the estimated failure probability. To address this limitation, GLHS introduces an iterative refinement procedure based on the identification of a buffer zone surrounding the estimated limit state. Within this region, additional $g_T$ evaluations are selectively introduced to build local surrogates, thereby improving the overall surrogate accuracy, while avoiding unnecessary computations elsewhere in the domain. The resulting framework naturally integrates surrogate modeling, domain learning (DL), and CS into a unified reliability analysis methodology.

In more detail, the GLHS methodology consists of the following four main steps, which are detailed in the remainder of this section:
\begin{enumerate}
    \item[Step 1)] Construction of a global surrogate $g_S$ over the input domain $\mathcal{D}$ using an initial set of $g_T$ evaluations. The global surrogate can be sampled at negligible cost to generate $m_h$ surrogate evaluations, providing an initial discrete representation of the domain and enabling the identification of an initial buffer zone. 
    See subsection \ref{subsec:init}.

    \item[Step 2)] Buffer zone domain learning and construction of an optimal sampling measure at iteration $l = 1,2,\dots$. At each iteration, a potential buffer zone is identified using the current surrogate $\tilde{g}^{(l-1)}$. To ensure a sufficiently rich discrete representation of the buffer zone, the samples are augmented to $m_d\gg 1$ by resampling within a hyperrectangle that encloses the buffer zone, consistent with the assumption of uniformly distributed inputs. An optimal sampling measure, i.e., CS, is then constructed on this discrete set to select a limited number of candidate points. 
    See subsection \ref{subsec:DL}.

    \item[Step 3)] Construction of a local surrogate $g_L^{(l)}$ at iteration $l = 1,2,\dots$ using the Christoffel samples generated in Step 2 in order to refine the accuracy of the surrogate near the limit state surface. See subsection \ref{subsec:lscons}.

    \item[Step 4)] Estimation of the probability of failure using the combined surrogate $\tilde{g}^{(l)}$, obtained by replacing the global surrogate with the local surrogate inside the buffer zone. See subsection \ref{subsectionPf}.
\end{enumerate}

Steps~2 and~3 are repeated until a prescribed stopping criterion is met, such as convergence of the estimated failure probability. While the GLHS framework is compatible with multiple surrogate reconstruction techniques, in the present work, a least-squares formulation is employed for both the global and local surrogate models.

\subsection{Global surrogate construction}
\label{subsec:init}

The first step of the GLHS framework consists of constructing global surrogate $g_S(\bm x)$ using realizations $\{g_T(\bm{x}_i)\}_{i=1}^{m_0}$ corresponding to samples  $\mathcal{S}_0=\{\bm{x}_i\}_{i=1}^{m_0}$ drawn from $\tau$ (in this work) or the CS strategy discussed in Section \ref{subsec:CS}. While the GLHS framework applies to arbitrary distributions $\tau$, we restrict our focus here to uniform distributions, i.e., $\bm{x}\sim\mathcal{U}[-1,1]^d$. The primary role of this surrogate is not to accurately resolve the failure boundary, but rather to enable efficient identification of regions near the limit state surface where local refinement is required. Because the $g_T$ evaluations are computationally expensive, $m_0$ is kept small.

Our approach involves utilizing a PCE to fit the obtained data  $\{ \bm{x}_i,g_T(\bm{x}_i)\}_{i=1}^{m_0}$. The PCE surrogate $g_S$ is the solution of a least squares regression problem. Consider a finite-dimensional space $P_\phi=\mathrm{span}\{\phi_1,\dots,\phi_N\}$, where $\{\phi_1,\dots,\phi_N\}$ represents a set of basis functions that are orthonormal with respect to $\tau$, here Legendre polynomials. The least squares approximation is defined as
\begin{equation}
    g_S := \argmin_{g_S\in P_\phi} \frac{1}{m_0}\sum_{i=1}^{m_0}|g_T(\bm{x}_i)-g_S(\bm{x}_i)|^2.
\end{equation}
Alternative PCE formulations, such as compressed sensing  \cite{doostan2011non,HAMPTON2015363}, may also be employed within the GLHS framework. In particular, $g_S$ may also be constructed using Christoffel sampling to define an optimal sampling measure for the global approximation, following the methodology described in Section~\ref{subsec:CS}. In the present implementation, Christoffel sampling is employed for the construction of the local surrogate, as described in Subsection~\ref{subsec:lscons}.

Once constructed, $g_S$ can be evaluated at negligible cost on a dense grid to reach $m_K\gg1$ realizations, enabling the identification of an initial buffer zone surrounding the estimated limit state surface. Specifically, a set of candidate samples is defined using a threshold $\eta_0>0$ that reflects the accuracy of the global surrogate near $g_T(\bm{x})=0$, and such that $|g_S(\bm{x})|\leq\eta_0$, as in \cite{LI20108966}. The buffer zone holds $m_h\le m_K$ samples $\mathcal{X}_1 = \{\bm{x}_i\}_{i=1}^{m_h}$.

The threshold $\eta_0$ is chosen based on the observed discrepancy between the global surrogate and $g_T$ evaluations:
\begin{equation}
\label{eq:eta_0Yset}
\eta_0
= c \max_{\bm{x}\in\mathcal{Y}} |g_T(\bm{x})-g_S(\bm{x})|,
\qquad
\mathcal{Y}
=\{\bm{x}_i\in\mathcal{S}_0:
|g_T(\bm{x}_i)| \le \alpha \max |g_T|\},
\end{equation}
where $c\ge 1$ controls the conservativeness of the buffer zone and
$\alpha\in(0,1]$ specifies the fraction of samples closest to the limit state surface. This construction ensures that the initial buffer zone is sufficiently large to contain the true failure boundary with high probability, while avoiding unnecessary refinement far from the limit state surface. The threshold $\eta_0$ is used only for the first iteration of GLHS.
Subsequent buffer zone thresholds are determined adaptively based on local surrogate accuracy, as described in Section~\ref{subsec:lscons}.

The work of \cite{LI20108966} discusses the difficulty of finding a conservative range for $\eta$. Ideally, $\eta$ should be small, but it cannot be too small as the limit state surface might be missed. The approach of \cite{LI20108966} relies on an approximate $\ell^{p}$-norm, $p\geq 1$, of the discrepancy between $g_T$ and $g_S$ to set $\eta$.

\subsection{Buffer zone domain learning and optimal sampling measure} 
\label{subsec:DL}

\begin{figure}
    \centering
    \includegraphics[width=0.75\linewidth]{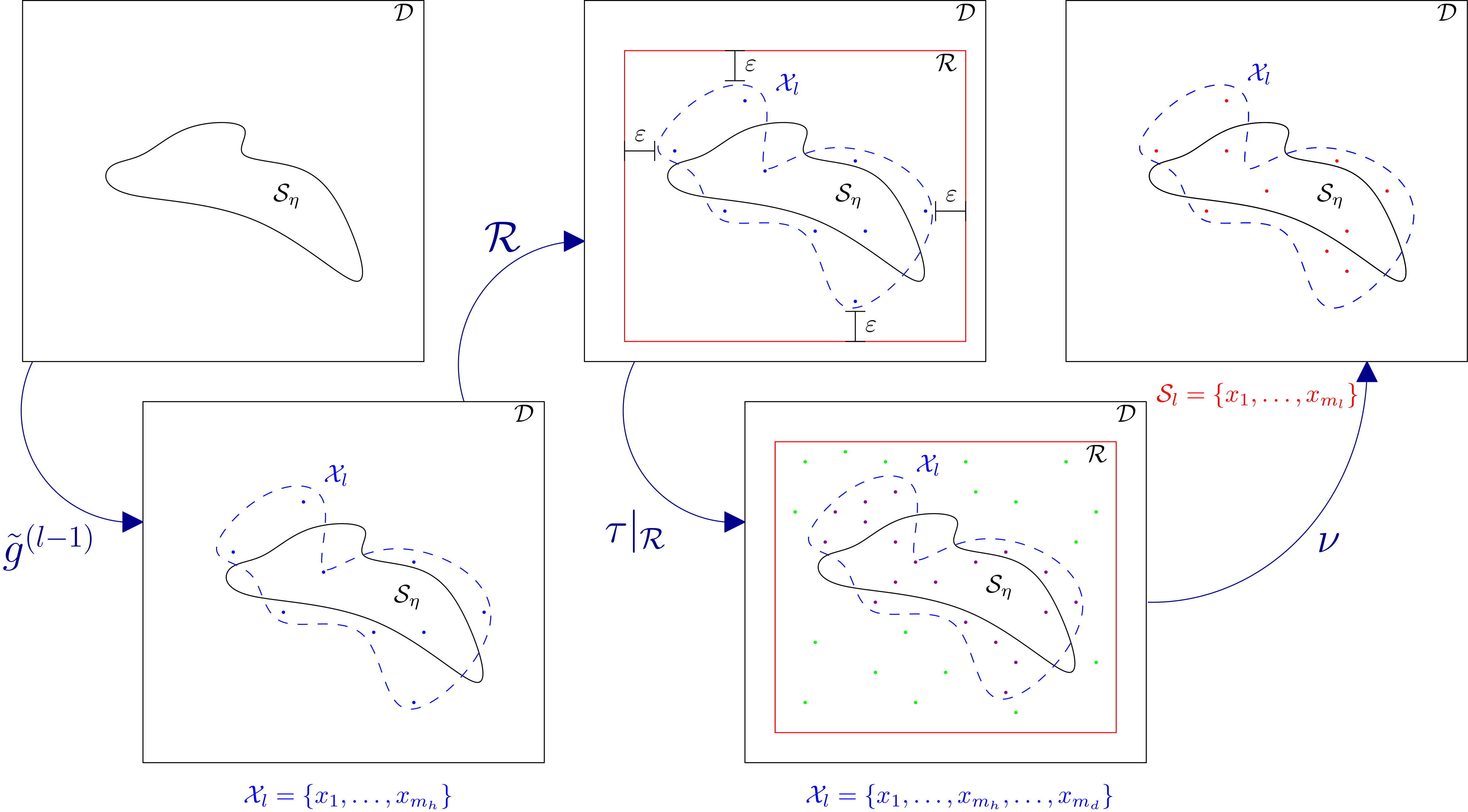}
    \caption{Visualization for the iterative procedure in buffer zone domain learning and optimal sampling measure. Starts with computing a buffer zone using the surrogate. Next, we construct the hyperrectangle to resample in the next step, and finally, we construct a sampling measure to draw new samples.}
    \label{fig:diagram-domain-learning-opt-measure}
\end{figure}

The second step augments the number of samples of the previously identified buffer zone to $m_d\gg1$ to learn the buffer zone domain. For the case of uniform $\tau$, and to accelerate the sampling, this is achieved by constructing a hyperrectangle that contains the estimated buffer zone. Next, we draw new samples by using an optimal sampling measure on the domain by CS, denoted as $\{\bm{x}_i\}_{i=1}^{m_l}$, which later will be used to construct the local surrogate $g_L$ over the estimated buffer zone. A diagram of the high-level description of this step is shown in Figure~\ref{fig:diagram-domain-learning-opt-measure}.

The method is based on General Domain Adaptive Strategy (GDAS) \citep{doi:10.1137/22M1472693}, where the main goal is to learn an unknown domain of interest within the input domain, here the buffer zone, of the function being approximated, while learning the target function from its samples. Throughout this section, we assume $g_T$ is a smooth, multivariate function. Given a threshold \(\eta > 0\), we define the buffer zone as 
\begin{equation}
    \mathcal{S}_\eta = \{\bm{x} \in \mathcal{D} :\left| g_{T}(\bm{x}) \right| \leq \eta \}.
\end{equation}
Since evaluations of the \(g_T(\bm{x})\) are computationally intensive,  characterizing the corresponding domain directly is not feasible. To address this, we adapt the GDAS methodology into a domain learning (DL) framework that uses the surrogate $g_S$ to approximate the buffer zone as
\begin{equation} 
\label{eq:Xl}
    \mathcal{X}_l = \{\bm{x}\in \mathcal{X}: |\tilde{g}^{(l-1)}(\bm{x})|\leq \eta_{l-1}\}, \quad l\geq 1,
\end{equation}
\textcolor{black}{where $\tilde{g}^{(l)}$ is the $l$-th surrogate of $g_T$ constructed, and $\eta_{l-1}$ is the threshold identified based on $\tilde{g}^{(l-1)}$, following the approach of Section~\ref{subsec:init}.}
Through $\tilde{g}^{(l-1)}$, we augment the samples in the grid $\mathcal{X}_l$ with the goal of resampling the buffer zone, making it rich for constructing the CS sampling on it. The Christoffel samples {$\{\bm{x}_i\}_{i=1}^{m_l}$} and the corresponding data {$\{g_T(\bm{x}_i)\}_{i=1}^{m_l}$} are then used to build the new local surrogate $g_L^{(l)}$.

In more detail, given the global grid $\mathcal{X}=\{\bm{x}_i\}_{i=1}^{m_K}\subset\mathcal{D}$, the subset
$\mathcal{X}_l\subseteq\mathcal{X}$ denotes the samples falling inside the current buffer zone. To accelerate drawing samples within the volume represented by $\mathcal{X}_l$, in the case of uniform $\tau$, we construct a hyperrectangle that encloses $\mathcal{X}_l$.  
Specifically, we consider an axis-aligned hyperrectangle obtained by computing, for each coordinate $j=1,\ldots,d$, the componentwise extrema
\[
a_j=\min_{\bm{x}_i\in\mathcal{X}_l}x_j^i-\varepsilon,
\qquad
b_j=\max_{\bm{x}_i\in\mathcal{X}_l}x_j^i+\varepsilon,
\]
where $\varepsilon>0$ is a small expansion parameter.  The resulting hyperrectangle is
\[
\mathcal{R}=[a_1,b_1]\times\cdots\times[a_d,b_d].
\]
The purpose of the threshold $\varepsilon>0$ is to extend the hyperrectangle, so that the volume of resampling covers a larger volume of the potential buffer zone. However, an unnecessarily large $\varepsilon$ will lead to an increased cost of building the local surrogate. Now, we resample within the hyperrectangle following $\tau$ restricted on the hyperrectangle. To do this, we draw a batch of $m_r > m_d$ samples from the hyperrectangle. We then evaluate whether each new sample belongs to $\mathcal{X}_1$ using the surrogate model $\tilde{g}^{(l-1)}$; samples that do not satisfy this condition are rejected. This process is repeated until $m_d\gg1$ valid samples in $\mathcal{X}_1$ are collected. Note that the computational cost remains low since the surrogate model is used to evaluate $g$ at the new samples.

Thus far, we have generated a dense grid $\mathcal{X}_l$ representing the approximate buffer zone and the discrete measure $\tau$ on this grid. We next proceed to build the CS measure and the local orthonormal basis $\{\phi_1,\dots,\phi_N\}$ associated with this buffer zone, following the strategy presented in Section \ref{subsec:CS}. These samples will be used to construct a local surrogate $g_L^{(l)}$ as explained next.

Further information concerning the algorithm implementation can be found in Appendix \ref{appendix:DL}.

\subsection{Local surrogate construction}
\label{subsec:lscons}

The third step of the GLHS framework consists of constructing a local surrogate by evaluating $g_T$ at the newly selected samples $\{\bm{x}_i\}_{i=1}^{m_l}$ within the estimated buffer zone. These evaluations yield the dataset $\{\bm{x}_i, g_T(\bm{x}_i)\}_{i=1}^{m_l}$, used to fit a polynomial surrogate in $\{\phi_1,\dots,\phi_N\}$ from Step 2 via least squares regression and cross-validation to optimally set the polynomial order. The resulting error metric, expressed in terms of the mean square error (MSE), is subsequently used to update the buffer zone threshold and determine whether additional refinement iterations, i.e., local surrogates, are required.

In detail, we seek to obtain $g_L^{(l)}$ such that 
\begin{equation}
    \label{def:regression-problem}
    g_L^{(l)}:= \argmin_{g_L^{(l)} \in P_\phi} \frac{1}{m_l}\sum_{i=1}^{m_l}w(\bm{x}_i)|g_T(\bm{x}_i)-g_L^{(l)}(\bm{x}_i)|^2,
\end{equation}
where $P_\phi=\mathrm{span}\{\phi_1,\ldots,\phi_N\}$. In \ref{def:regression-problem}, the weights $w(\bm{x}_i)$ are defined as in Section \ref{subsec:CS} and introduced as inputs $\bm x$ are sampled from the CS measure instead of $\tau$; see \cite{adcock2019nearoptimalsamplingstrategiesmultivariate}. Writing $g_L^{(l)}=\sum_{j=1}^Nc_j\phi_j(\bm{x})$, we seek to approximate $\bm{c}=(c_i)_{i=1}^{N}$ such that 
\begin{equation}
\label{eqn:ls}
    \hat{\bm{c}} :=\argmin_{\tilde{\bm{c}}\in\mathbb{R}^N}\|\bm{A}\tilde{\bm{c}}-\bm{b}\|_{2}.
\end{equation}
In the weighted least squares formulation, $\bm{A}\in \mathbb{R}^{m_l \times N}$ is the measurement matrix of the basis functions evaluated at the sample, and $\bm{b} \in \mathbb{R}^{m_l}$ is the vector of target function values (e.g., $g_T(\bm{x})$) at the corresponding sample. Specifically,  $\bm{A} $ and $\bm{b}$ are given by
\begin{equation}
    \bm{A} = \bigg\{ \sqrt{\frac{w(\bm{x}_i)}{m_l}}\phi_j(\bm{x}_i) \bigg\}^{m_l,N}_{i=1,j=1}, \quad  \bm{b} = \bigg\{ \sqrt{\frac{w(\bm{x}_i)}{m_l}}g_T(\bm{x}_i) \bigg\}^{m_l}_{i=1}.
\end{equation}
Following \cite{Hampton_2015,adcock2019nearoptimalsamplingstrategiesmultivariate}, the least squares formulation (\ref{eqn:ls}) leads to stable solutions as long as $m_l$ follows a sampling rate
\begin{equation}
\label{eqn:sampling_rate}
   m_l \sim N \log(N),  
\end{equation}
or empirically $m_l\approx 4N$. 

Recall that $g_L^{(l)}$ is a polynomial expansion of some order $n$ and that the Christoffel sampling measure $\nu$ depends on the choice of the polynomial space, hence $n$. For the ease of implementation, in our numerical experiments, we set $n = n_{\max}=3$ to generate the Christoffel samples, but use cross-validation to determine the optimal order \(n\) for each  $g_L^{(l)}$ construction. This is done to keep the number $m_l$ evaluation of $g_T$, given the sampling rate (\ref{eqn:sampling_rate}), and that $N$ grows as a function of $n$. Starting from \(n_{\max}\), we consider 
\(1 \leq n \leq n_{\max}\) basis and construct a local surrogate for each $n$. For these fits, we specify the number of samples $m_l$ based on (\ref{eqn:sampling_rate}).

Note that the local surrogate $g_L^{(l)}$ is a polynomial expansion of total order $n$, and that the Christoffel sampling measure $\nu$ depends on the choice of the polynomial space, and hence on $n$. In the present implementation, for ease of use, we fix a maximum order $n_{\max}$ and construct the Christoffel sampling measure using the basis associated with $n_{\max}$. A total of $m_l$ candidate samples is then drawn according to this measure, where $m_l$ is chosen to satisfy the sampling rate requirement \eqref{eqn:sampling_rate} for $n_{\max}$. To determine the appropriate local model complexity, we subsequently restrict the polynomial space to orders $1 \leq n \leq n_{\max}$ and construct a surrogate for each $n$ using the same candidate set, with the number of samples $m_l$ divided in a training and verification set. Cross-validation is then employed to select the polynomial order that minimizes the prediction error. This strategy ensures that the total number of $g_T$ evaluations remains controlled while allowing flexibility in the surrogate order.

While the above approach evaluates $g_T$ at all $m_l$ samples associated with $n_{\max}$, a more efficient procedure can be devised by exploiting the nested structure of polynomial spaces. Specifically, one may first generate a candidate set of size $m_l(n_{\max})$ using the Christoffel measure for $n_{\max}$, but initially evaluate the $g_T$ only at the subset $m_l(1)$ (obtain using \eqref{eqn:sampling_rate} and $n=1$), required for a first-order surrogate. The local surrogate is then constructed and its error assessed. If the error exceeds a prescribed tolerance, the polynomial order is increased to $n=2$, and only the additional samples needed to reach $m_l(2)$ are evaluated. This process is repeated for $n=3,\ldots,n_{\max}$ until the error criterion is satisfied or $n_{\max}$ is reached. In this manner, $g_T$ evaluations are introduced incrementally and only as required, thereby minimizing computational cost while retaining the stability guarantees of Christoffel sampling.

\begin{figure}[h]
\centering
\includegraphics[width=0.85\textwidth]{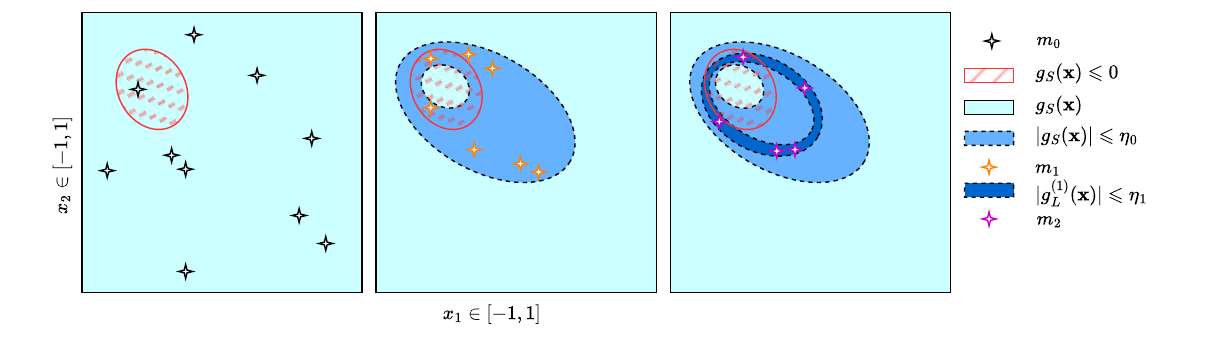}
\caption{The first box represents the method described in the initialization \ref{subsec:init}. The second is the representation of the procedure that yields the first local surrogate $g_L^{(1)}(\bm{x})$ as described in \ref{subsec:lscons}. The last box represents the procedure to obtain the second local surrogate $g_L^{(2)}(\bm{x})$. }
\label{2dillust}
\end{figure}

\paragraph{Updating global surrogate and computing $\eta_l$.} Given the local surrogate $g_L^{(l)}$, we update the overall (global) surrogate $\tilde{g}^{(l)}$ and estimate the threshold $\eta_l$ for the identification of the new buffer zone associated with $\tilde{g}^{(l)}$. Specifically, we set 
\begin{equation}
\tilde{g}^{(l)}(\bm{x}) =
\begin{cases}
g_{L}^{(l)}(\bm{x}), & \bm{x}\in\mathcal{X}_0 \ :\ |\tilde{g}^{(l-1)}(\bm{x})| \le \eta_{l-1},\\
\tilde{g}^{(l-1)}(\bm{x}), & \bm{x}\in\mathcal{X}_0 \ :\ |\tilde{g}^{(l-1)}(\bm{x})| > \eta_{l-1}.
\end{cases}
\end{equation}
and 
\begin{equation}
    \eta_l 
    = \left(\sum_{i=1}^{m_l} |g_T(\bm{x}_i)-\tilde{g}^{(l)}(\bm{x}_i)|^2 \right)^{1/2}.   
\end{equation}
We note that our choice of $\eta_l$ is based on the MSE of the updated surrogate, which is different from how $\eta_0$ was determined. Our empirical results illustrated the superiority of MSE, which we explained by the relatively small volume of the subsequent buffer zone. The process of updating the global surrogate and determining the new buffer zone is illustrated in Figure \ref{2dillust}.

\paragraph{Remark:} In the present study, the regression data $\{\bm{x}_i, g_T(\bm{x}_i)\}_{i=1}^{m_l}$ at iteration $l$ of GLHS does not include data from iteration $l-1$, for the interest of simplifying the construction of Christoffel samples. Future work may consider nested sampling strategies, such as that of \cite{HAMPTON201820}, to incorporate those data and, hence, further reduce the number of $g_T$ evaluations.

\paragraph{Convergence of the GLHS iterations.} Assuming that the estimated buffer zone identified via the surrogate model at each iteration of GLHS (see \ref{eq:Xl}) contains the limit state surface and $g_T$ varies smoothly in the vicinity of the limit state surface, the local surrogate improve the approximation of $g_T$. This implies that the threshold parameter $\eta_{l}$, if accurately estimated, becomes progressively smaller, the subsequent buffer zones will have smaller volume; thus, the error in estimating $P_f$ becomes smaller. In applying GLHS to various test cases, it was observed that iteration $l=1$ already produced a significant improvement in the estimation of the probability of failures. In addition, for the test cases analyzed, the local surrogate resulted in a small value for $\eta_1$, leading to a limited number of samples in the buffer zone, perhaps, indicating the convergence of the method. In such cases, to perform iteration 2, either the number of initial samples $m_K$ has to be increased or the value of $\eta_1$ has to be increased.

\begin{figure}[h]
    \centering
    \begin{subfigure}[t]{0.285\textwidth}
        \centering
        \includegraphics[width=\textwidth]{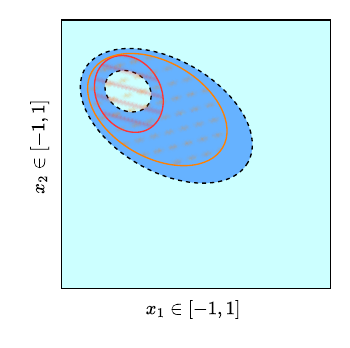}
        \caption{}
        \label{fig:afailure}
    \end{subfigure}
    \begin{subfigure}[t]{0.5\textwidth}
        \centering
        \includegraphics[width=\textwidth]{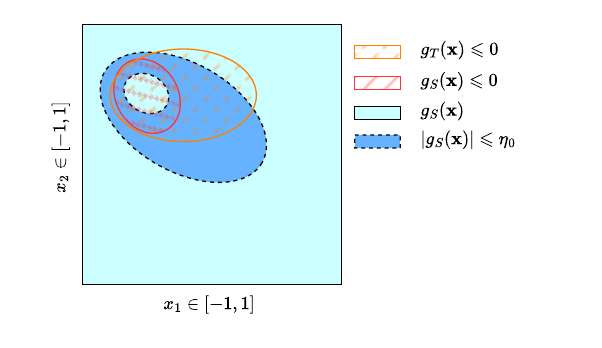}
        \caption{}
        \label{fig:bfailure}
    \end{subfigure}
    \caption{Figure \ref{fig:afailure} is the representation for the optimum situation: the buffer zone $|g_S(\bm{x})|\leq 0$ fully encompasses the limit state surface, which is represented by the solid orange line. Figure \ref{fig:bfailure} represents the case where the buffer zone was not large enough and thus $g_T=0$ is only partially captured. In this situation, a bias will exist in the probability of failure estimation as the samples not captured by the buffer zone will not get updated.}
\label{fig:abfailure}    
\end{figure}

\begin{figure}[h]
    \centering
    \begin{subfigure}[t]{0.285\textwidth}
        \centering
        \includegraphics[width=\textwidth]{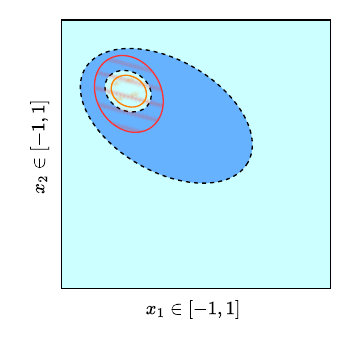}
        \caption{}
        \label{fig:cfailure}
    \end{subfigure}
    \begin{subfigure}[t]{0.5\textwidth}
        \centering
        \includegraphics[width=\textwidth]{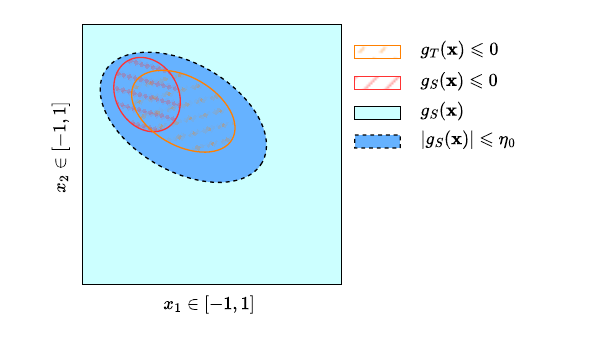}
        \caption{}
        \label{fig:dfailure}
    \end{subfigure}
    \caption{Figure \ref{fig:cfailure} represents the case where the buffer zone is not large enough and thus $g_T=0$ is not captured. In this situation, a bias will exist in the probability of failure estimation. Figure \ref{fig:dfailure} represents a scenario where the entire domain of failure is being updated by the local surrogate.}
\label{fig:cdfailure}
\end{figure}

In our test cases, we also observed that the estimated failure probability can be biased if the identified buffer zone at iteration 1 is not large enough due to underestimating $\eta_0$. Four scenarios related to this are illustrated in Figures \ref{fig:abfailure} and \ref{fig:cdfailure}. 

\subsection{Estimation of probability of failure}
\label{subsectionPf}
In this work, we define the estimated probability of failure as follows
\begin{equation}
    P_{f}^{GLHS}\  = \frac{1}{m_{c}}\ \sum_{i = 1}^{m_{c}}{\mathbb{I}_{\tilde{g}^{(l)}\leq 0}(\bm{x}_i)},
\end{equation}
where \(m_{c}\) denotes the total number of Monte Carlo samples evaluating the surrogates $\tilde{g}^{(l)}$ at small computational cost. 

In the examples provided in the next section, the value of \(m_{c}\) is set to $m_c =10^6$, although it can be adjusted according to the anticipated ranges of target failure probability.

\section{Numerical Experiments}
\label{sec:ne}
\subsection{One-dimensional (1D) test case}
We begin with a one-dimensional test case to illustrate the main components of the GLHS methodology in a setting where the behavior of the limit-state function and the impact of buffer zone refinement can be clearly visualized. Although low-dimensional, this example is intentionally chosen to expose both the strengths of the method and its sensitivity to the initial buffer zone construction.

The input variable $x$ is assumed to be uniformly distributed over the interval $[-1,1]$. The ground-truth limit-state function $g_T$ is defined as
\begin{equation}
    g_T ({x}) = \left(-\frac{10}{16}{x}^4 + \frac{15}{16}{x}^3-\frac{15}{8}{x}^2-\frac{4}{16}{x}+5\right)e^{-{x}} -2.
\end{equation}
Failure is defined by the condition $g_T(x) \le 0$. Using the Monte Carlo estimator in Eq.~\ref{eq:PFMC}, with $g_T$ evaluated on $m_c$ samples, the failure domain is found to occupy approximately $15\%$ of the input space. This corresponds to a relatively high-probability failure scenario and provides a baseline test case in which surrogate-induced bias in the failure probability estimate is expected to be readily observable.

The algorithm parameters used in this example are summarized in Table~\ref{Table1Dinputs}. A global surrogate $g_S$ is constructed using $m_0 = 5$ samples and a low polynomial order $n = 2$. The surrogate is evaluated on the $m_c$ Monte Carlo samples for failure probability estimation and on a smaller set of $m_K$ samples to support the GLHS buffer zone identification and refinement.  

\begin{table}[htbp]
\centering
\begin{tabular}{l l c} 
 \hline
  Parameter & Description & Value \\ 
 \hline
 $m_K$ & Number of samples in the domain & $10000$ \\ 
 $m_{c}$ & Number of MC samples & $10000000$ \\ 
 $d$ & Number of dimensions & 1 \\ 
 $c$ & Constant controlling the buffer zone volume & 1 \\
 $\alpha$ & \% of data used to create initial buffer zone & 0.8 \\
 $m_{0}$ & Number of initial samples to compute $g_S$ & 5 \\ 
 $n$ & Order of global surrogate & 2 \\
 $n_{max}$ & Maximum order of local surrogate & 3 \\
 $m_{l}$ & Number of samples to compute $g_L^{(1)}$ & 6 \\
 $c_{r}$ & Constant to proportionally add samples in the buffer zone & 1.5 \\
 \hline
\end{tabular}
\caption{Initial parameters set by the user for the 1D test case.}
\label{Table1Dinputs}
\end{table}

\begin{figure}[htbp]
    \centering
    \begin{subfigure}[t]{0.45\textwidth}
        \centering
        \includegraphics[width=\textwidth]{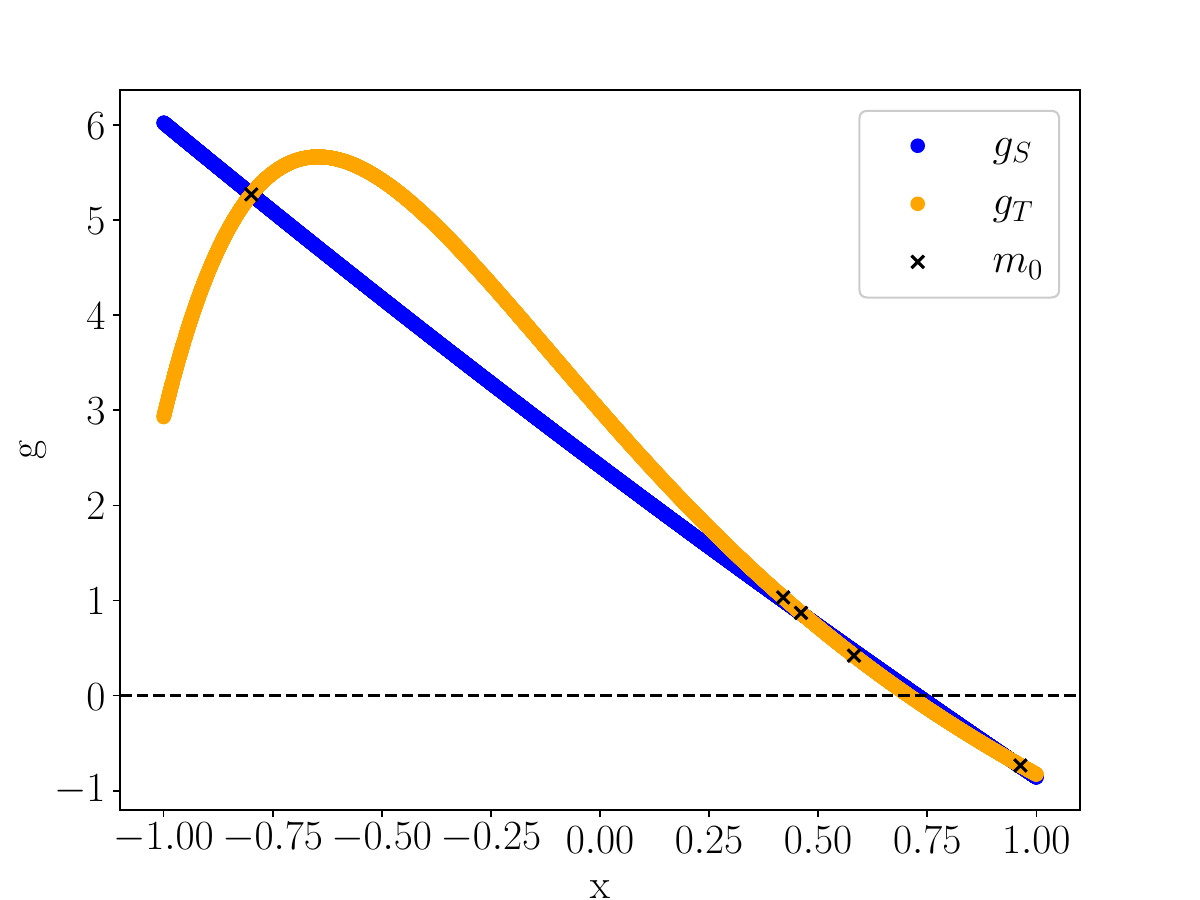}
        \caption{}
        \label{fig:1D_S_T}
    \end{subfigure}
    \hspace{0.05\textwidth}
    \begin{subfigure}[t]{0.45\textwidth}
        \centering
        \includegraphics[width=\textwidth]{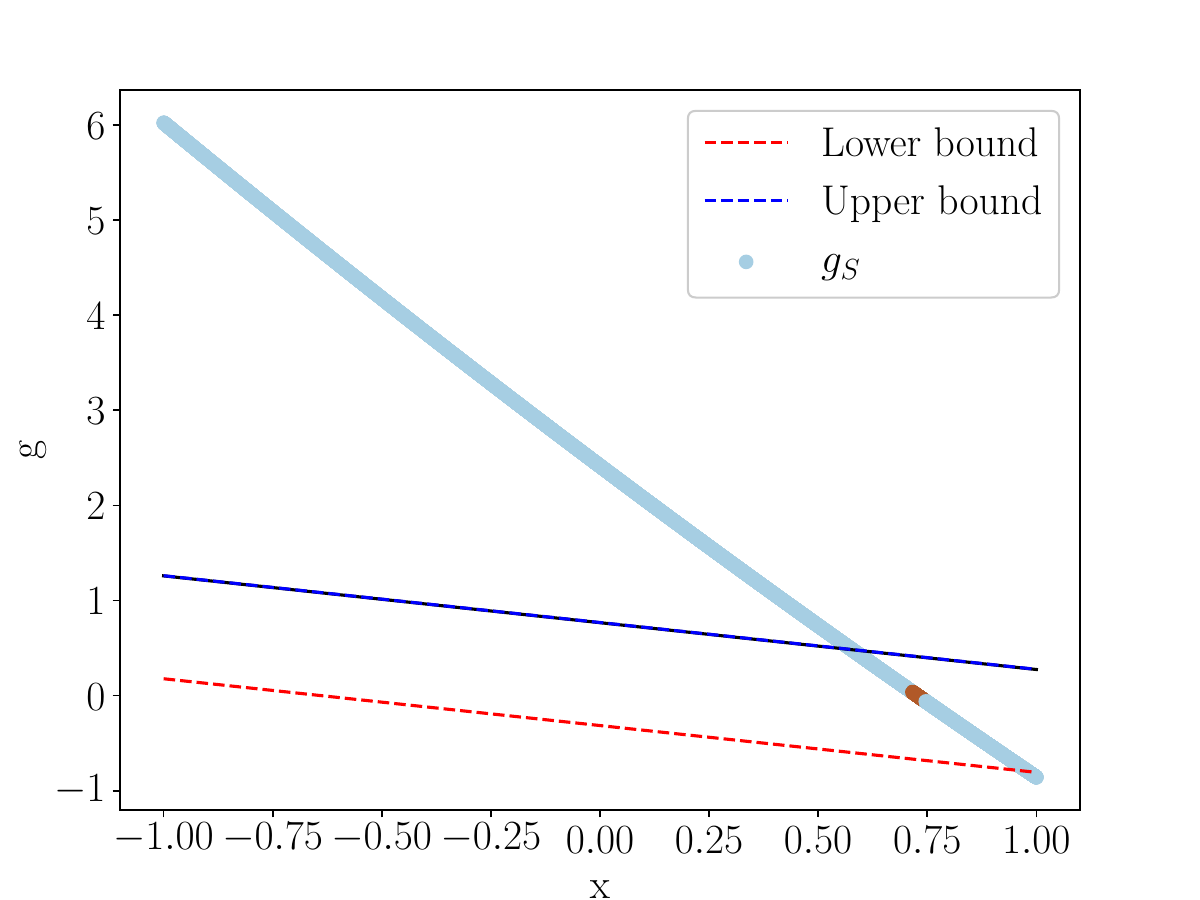}
        \caption{}
        \label{fig:1D_hyper}
    \end{subfigure}
    \caption{Figure \ref{fig:1D_S_T} shows the ground truth function $g_T$ and the global surrogate $g_S$ for the 1D test case. The initial $m_0$ samples used to construct the global surrogate are marked with crosses. Figure \ref{fig:1D_hyper}  illustrates the hyperplanes used to define the bounds for selecting the region of interest on the surrogate during resampling. The samples of $g_S$ that fall within the buffer zone are highlighted in orange.}
\end{figure}

Figure~\ref{fig:1D_S_T} compares $g_T$ and $g_S$, together with the $m_0$ samples used to construct $g_S$. Although the global surrogate exhibits noticeable discrepancies away from the limit-state surface, the zero-crossing of $g_T$ is reasonably well captured. As a result, using $g_S$ alone leads to a relative error of $6.8\%$ in the estimated probability of failure.

The buffer zone is identified using the initial threshold $\eta_0$ defined in Eq.~\ref{eq:eta_0Yset}, with the conservativeness parameter set to $c = 1$. In the one-dimensional setting, the associated hyperrectangle domain-learning step simplifies to defining two bounding hyperplanes that constrain the buffer zone to an interval in $x$, expanded by a small safety margin.

\begin{figure}[htbp]
    \centering
    \begin{subfigure}[t]{0.45\textwidth}
        \centering
        \includegraphics[width=\textwidth]{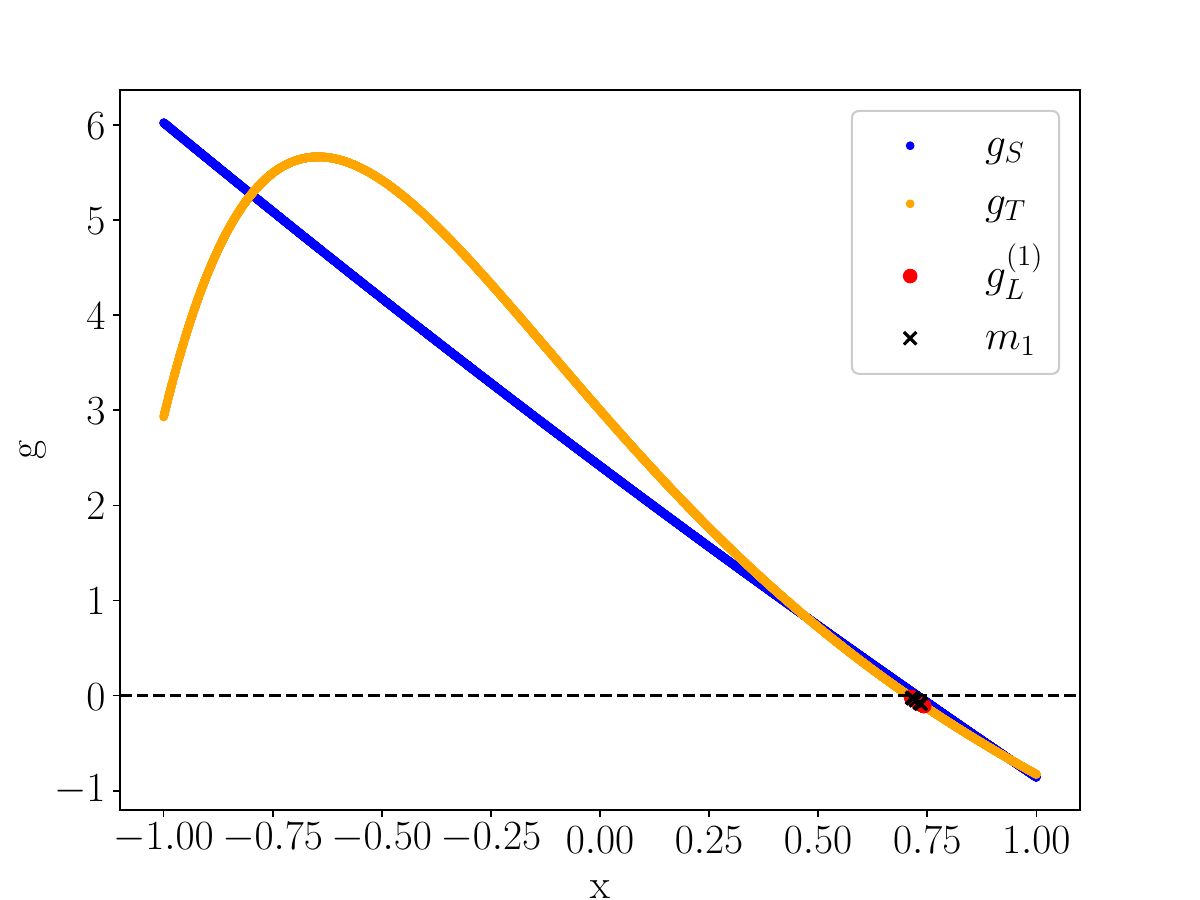}
        \caption{}
        \label{fig:1D_L_T}
    \end{subfigure}
    \hspace{0.05\textwidth}
    \begin{subfigure}[t]{0.45\textwidth}
        \centering
        \includegraphics[width=\textwidth]{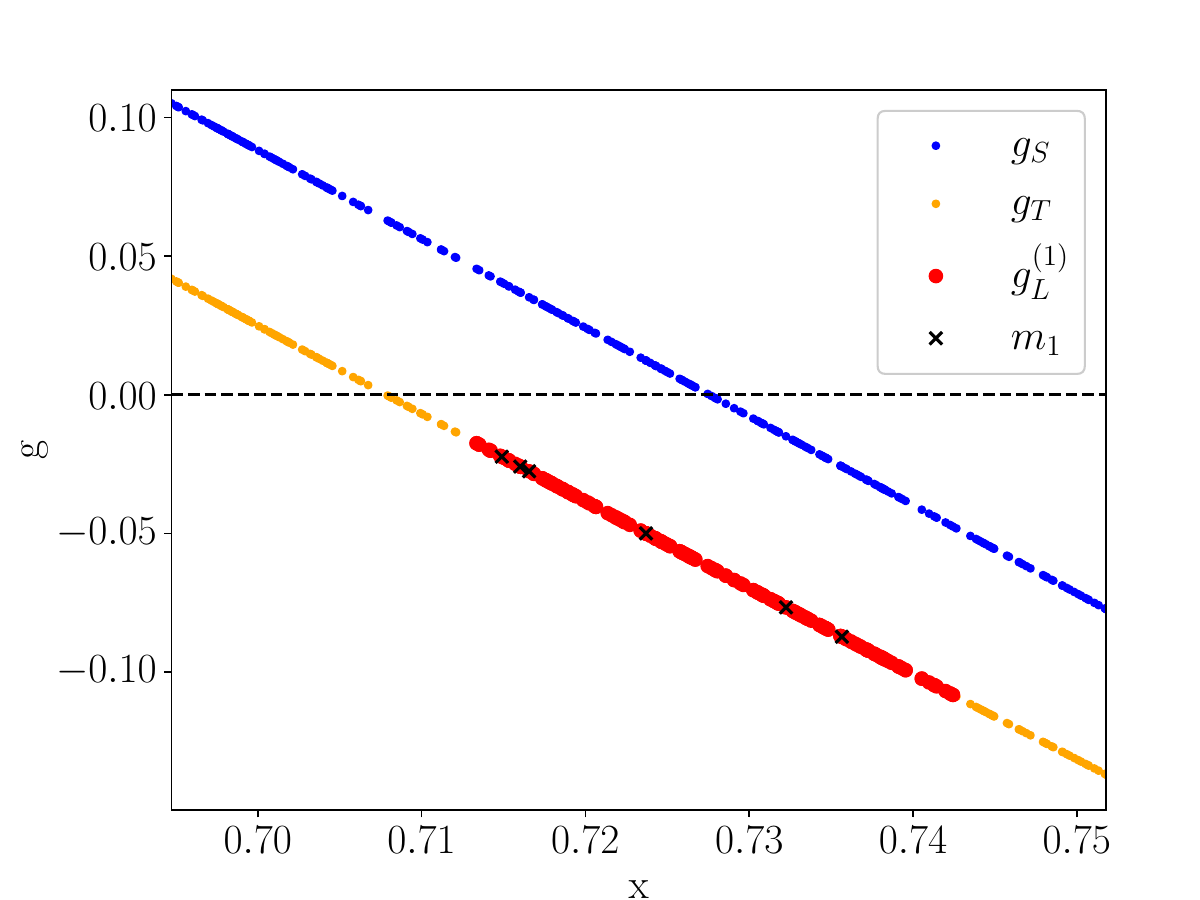}
        \caption{}
        \label{fig:1D_L_Tb}
    \end{subfigure}
    \caption{Figure \ref{fig:1D_L_T} shows the ground truth function $g_T$ and global surrogate $g_S$ for the 1D test case. The local surrogate is constructed using the $m_1$ samples selected by the domain learning strategy. In this example, the buffer zone—where the local surrogate $g_L^{(1)}$ is defined—contains 152 points. Figure \ref{fig:1D_L_Tb} provides a close-up view of the buffer zone from Figure \ref{fig:1D_L_T}.}
\end{figure}

Using Christoffel adaptive sampling within this interval, $m_1 = 6$ additional samples are selected to construct a local surrogate $g_L^{(1)}$. Cross-validation selects a polynomial order of $2$ for the local approximation. Figure~\ref{fig:1D_L_T} shows the resulting combined surrogate $\tilde{g}^{(1)}$, while Figure~\ref{fig:1D_L_Tb} provides a zoomed view of the buffer zone. The local surrogate closely matches the ground truth. As a result, the algorithm converges after the first iteration ($l = 1$), with no further refinement required since the updated threshold $\eta_1$ yields an empty buffer zone on $m_K$ samples.

Table~\ref{Table1DPf} reports the estimated probabilities of failure. All estimates are computed using the same Monte Carlo sample set of size $m_c$. Within the buffer zone, values of the global surrogate $g_S$ are replaced by the local surrogate $g_L^{(1)}$ to yield $\tilde{g}^{(1)}$, reducing the relative error in the failure probability estimate from $6.8\%$ to $1.6\%$ using only $m_T = m_0 + m_1 = 11$ model evaluations.

\begin{table} [htbp]
\centering
\begin{tabular}{l c c c c} 
 \hline
  Model & $\mu_{P_f}$ & $\%$ error with $g_T$ & Order local & $m_{T}$\\ 
 \hline
 \(g_{T}\) & 0.1462 & - & - & - \\ 
 \(g_{S}\) & 0.1363 & $6.8\%$ & - & 5 \\ 
 \({\tilde{g}}^{(1)}\) & 0.1439 & $1.6\%$ & 2 & $5+6$ \\  
 \hline
\end{tabular}
\caption{Results for the 1D test case, where the probability of failure \(P_f\) is estimated using Monte Carlo sampling. The quantity \(m_{\text{hf}}\) denotes the number of simulations required to obtain the reported results. This test case exhibits a relatively high
probability of failure, with $14.62\%$ of the Monte Carlo samples falling in the failure domain. The column 'order local' indicates the polynomial order selected by the algorithm for constructing the local surrogate. The algorithm was repeated 10 times, and the reported \(\mu_{P_f}\) corresponds to the mean estimate across these repetitions. The standard deviation was zero for the surrogate \(\tilde{g}^{(1)}\).}
\label{Table1DPf}
\end{table}

Although the improvement is substantial, Figure~\ref{fig:1D_L_Tb} shows that the initial buffer zone does not fully enclose the transition of the limit-state function $g_T$, leading to a small residual bias. This behavior is consistent with the discussion in Section~\ref{subsec:lscons} regarding insufficiently conservative buffer zone selection when estimating $\eta_0$.

\begin{figure}[htbp]
\centering
\includegraphics[width=0.45\textwidth]{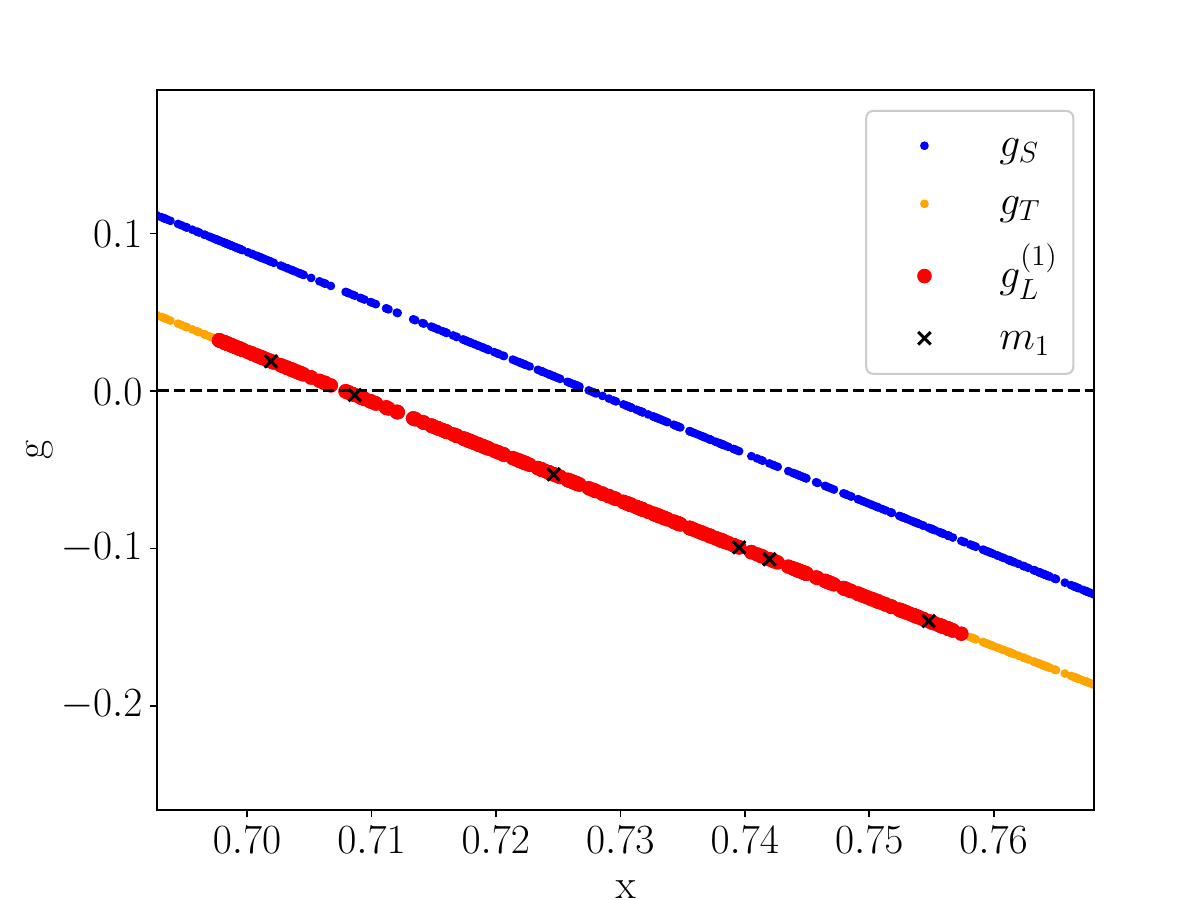}
\caption{This figure shows a zoom of the buffer zone for the 1D test case with increased conservativeness, using $c=2.0$.}
\label{1D_L_Tby}
\end{figure}

Increasing the conservativeness parameter to $c = 2.0$ enlarges the initial buffer zone. As shown in Table~\ref{Table1DPfy} and Figure~\ref{1D_L_Tby}, the local surrogate then fully encloses the transition region, leading to a zero relative error in the estimated failure probability.

\begin{table}[htbp]
\centering 
\begin{tabular}{l c c c c} 
 \hline
 Model & $\mu_{P_f}$ & $\%$ error with $g_T$ & Order local & $m_{\text{hf}}$\\ 
 \hline
 \(g_{T}\) & 0.1462 & - & - & - \\ 
 \(g_{S}\) & 0.1363 & $6.8\%$ & - & 5 \\ 
 \(\tilde{g}^{(1)}\) & 0.1462 & $0.0\%$ & 2 & $5+6$ \\  
 \hline
\end{tabular}
\caption{Results for 1D with probability of failure estimation through a Monte Carlo sampling. The buffer zone was updated here to $c=2.0$. The $\mu_{P_f}$ is the mean value of the algorithm being run 10 times. The standard deviation of $\tilde{g}^{(1)}$ in this case is $3.5e-7$.}
\label{Table1DPfy}
\end{table}

This test case was specifically chosen to illustrate the challenges associated with estimating the initial buffer zone threshold $\eta_0$ and the impact of its conservativeness on failure probability estimation. Overall, this example demonstrates that even a low-order global surrogate can provide a useful initial approximation, while the introduction of a local surrogate over an appropriately chosen buffer zone yields accurate failure probability estimates with a minimal number of simulations.

\subsection{Two-dimensional (2D) test case}

For this test case, we consider a two-dimensional input $\bm{x}$ uniformly distributed over the domain $[-1,1]^2$. The numerical investigation follows the same methodology and comparison strategy as in the one-dimensional test case.

The ground-truth limit-state function is defined as
\begin{equation}
    g_T(x_1,x_2) = x_1^4 \cos(x_1) + 0.5x_1^3 + 0.5x_1^2 + 0.5x_1x_2 - 0.5x_2^2 \cos(x_2) + 1.0.
\end{equation}
The algorithm parameters used in this example are summarized in Table~\ref{Table2Dinputs}.

\begin{table}[h]
\centering 
\begin{tabular}{l l c} 
 \hline
  Parameter & Description & Value \\ 
 \hline
 $m_K$ & Number of samples in the domain & $50000$ \\ 
 $m_{c}$ & Number of MC samples & $10000000$ \\ 
 $d$ & Number of dimensions & 2 \\ 
 $c$ & Constant controlling the buffer zone volume & 1 \\
 $\alpha$ & \% of data used to create initial buffer zone & 0.5 \\
 $m_{0}$ & Number of initial samples to compute $g_S$ & 40 \\ 
 $n$ & Order of global surrogate & 4 \\
 $n_{max}$ & Maximum order of local surrogate & 3 \\
 $m_{l}$ & Number of samples to compute $g_L^{(1)}$ & 17 \\
 $c_{r}$ & Constant to proportionally add samples in the buffer zone & 1.5 \\
 \hline
\end{tabular}
\caption{Initial parameters set by the user for the 2D test case.}
\label{Table2Dinputs}
\end{table}

\begin{figure}[htbp]
\centering
\includegraphics[width=1.0\textwidth]{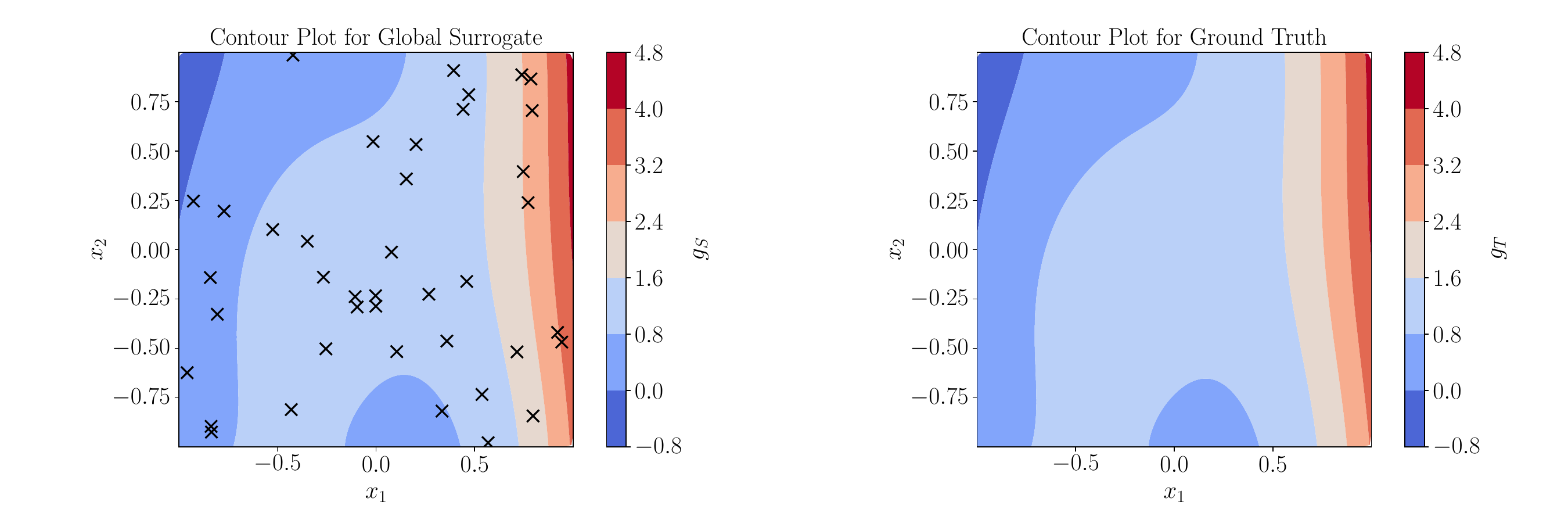}
\caption{Global surrogate and ground truth representation for the 2D test case. The samples $m_{0}$ used to compute the global surrogate are represented by a cross on the left plot.}
\label{2D_S_T}
\end{figure}

Figure~\ref{2D_S_T} shows contour plots of the ground-truth function $g_T$ and the global surrogate $g_S$, together with the $m_0$ samples used to construct the surrogate. In this two-dimensional case, the global surrogate provides a relatively accurate approximation of the limit-state surface. Based on $m_c$ Monte Carlo samples, the failure domain occupies approximately $2.54\%$ of the input space, corresponding to a smaller failure probability than in the one-dimensional example. Using $g_S$ alone results in a relative error of $3.7\%$ in the estimated probability of failure, which we aim to further reduce using the GLHS framework.

\begin{figure}[htbp]
    \centering
    \begin{subfigure}[t]{0.45\textwidth}
        \centering
        \includegraphics[width=\textwidth]{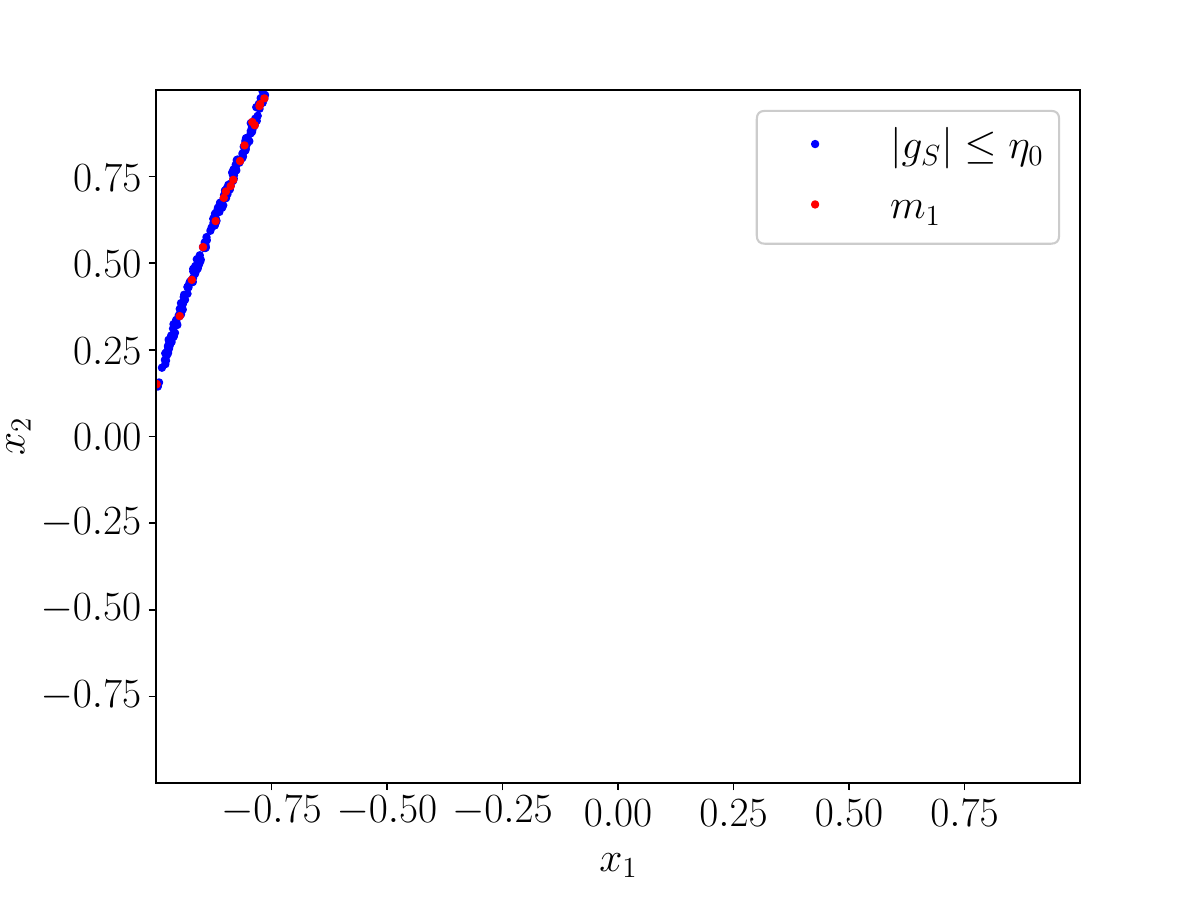}
        \caption{}
        \label{fig:2D_stratpts}
    \end{subfigure}
    \hspace{0.05\textwidth}
    \begin{subfigure}[t]{0.45\textwidth}
        \centering
        \includegraphics[width=\textwidth]{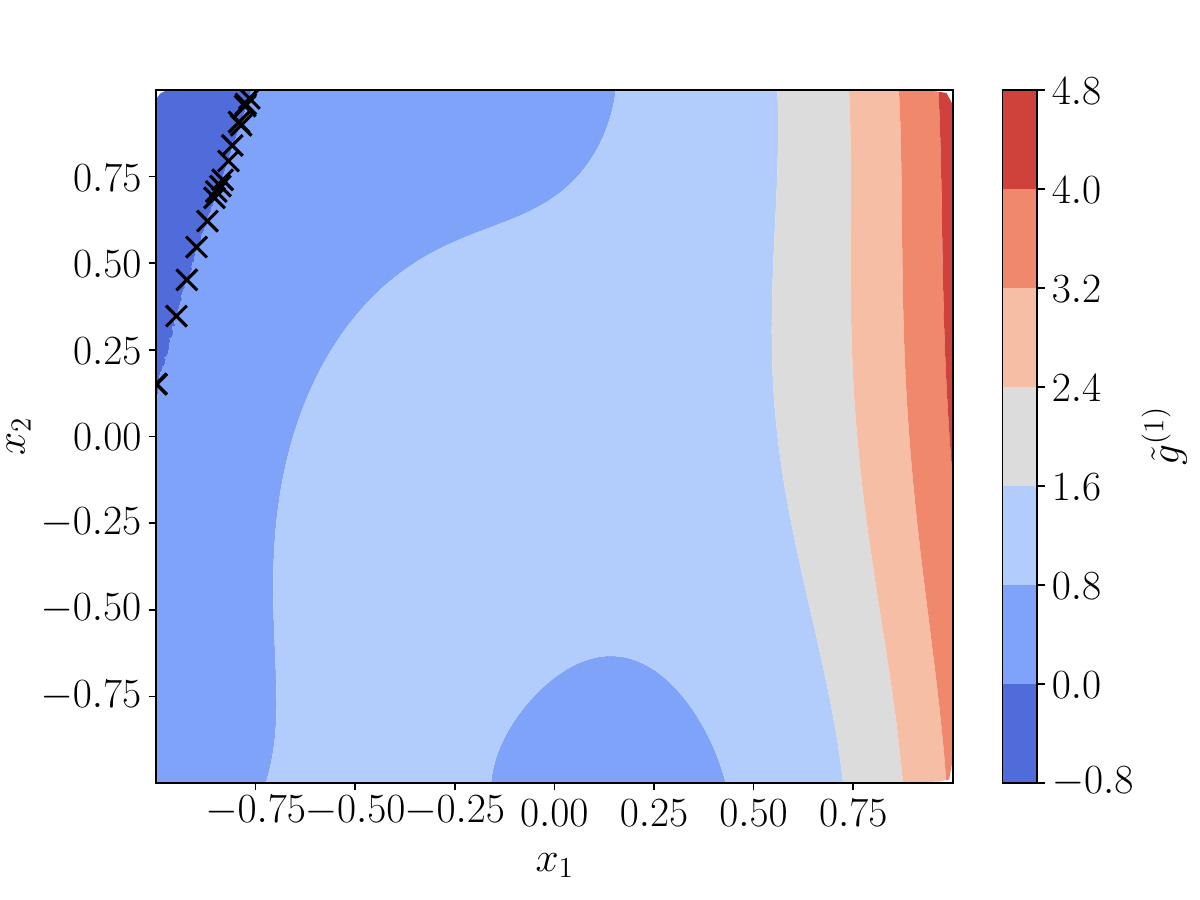}
        \caption{}
        \label{fig:2D_L_T}
    \end{subfigure}
    \caption{Figure \ref{fig:2D_stratpts} represents the buffer zone and the $m_1$ samples computed by the Domain Learning. Figure \ref{fig:2D_L_T} represents the final surrogate $\tilde{g}^(1)$ contour plot with the $m_1$ samples.}
\end{figure}

Figure~\ref{fig:2D_stratpts} illustrates the identified buffer zone and the corresponding domain-learning samples that form its support. In this test case, the buffer zone contains 123 out of the $m_K$ samples in the domain. The resulting combined surrogate is shown in Figure~\ref{fig:2D_L_T}, together with the $m_1$ samples selected by CS to construct the local surrogate.

The resulting failure probability estimates are reported in Table~\ref{Table2Dpf}. Constructing the global surrogate requires 40 model evaluations, while an additional 17 simulations are used to build the local surrogate. In this two-dimensional test case, the global surrogate alone yields a relative error of $3.5\%$ in the estimated probability of failure, which is reduced to $0.4\%$ when combined with a local surrogate. 

\begin{table}[htbp]
\centering
\begin{tabular}{l c c c c} 
 \hline
  Model & $\mu_{P_f}$ & $\%$ error with $g_T$ & Order local & $m_{T}$ \\ 
 \hline
 \(g_{T}\) & 0.0254 & - & - & - \\ 
 \(g_{S}\) & 0.0245 & $3.5\%$ & - & 40 \\ 
 \({\tilde{g}}^{(1)}\) & 0.0253 & $0.4\%$ & 2 & $40+17$ \\ 
 \hline
\end{tabular}
\caption{Results for 2D with probability of failure estimation through a Monte Carlo sampling. $10$ million samples were used in all 3 cases. The $\mu_{P_f}$ estimation of $\tilde{g}^{(1)}$ is a mean value from repeating the algorithm 10 times. The standard deviation in this case is $1.6e-4$.}
\label{Table2Dpf}
\end{table}

Next we compare the performance of the GLHS method against the non-iterative approach of \cite{LI20108966, LI20118683}, described in Section~\ref{subsec:NIM}, with results summarized in Table~\ref{Table2D_Li}. For this test case, the buffer zone identified by the global surrogate contains 23{,}133 samples. While the non-iterative method converges to the ground-truth failure probability, it does so at the expense of evaluating the model at all samples within the buffer zone. In contrast, the current GLHS implementation requires only 17 simulations per iteration to construct the local surrogate. To ensure a fair comparison, the non-iterative method is therefore restricted to the same number of evaluations, allowing assessment of the accuracy achievable under comparable computational budgets.

The non-iterative method is implemented by sampling the input uniformly over $[-1,1]^2$ and evaluating the surrogate $g_S(\bm{x})$ at each sample. Samples satisfying $|g_S(\bm{x})| \le \eta$ are retained until $m_1$ are collected, after which the failure probability is estimated following the procedure described in Section~\ref{subsec:NIM}. Specifically,
\[
\text{count} = 
\begin{cases}
\text{count} + 1, & \text{if } |g_S(\bm{x})| \le \eta,\\[6pt]
\text{count}, & \text{otherwise}.
\end{cases}
\]
Sampling continues until $\text{count} = m_1$, at which point the algorithm terminates.

For the purpose of comparison, a second iteration is also considered. In practice, the buffer zone often contracts rapidly; therefore, to ensure the availability of samples in the second iteration, the threshold is updated using
\begin{equation}
    \tilde{\eta}_1 = \frac{\eta_0 + \eta_1}{2} \ge \min\{\eta_0,\eta_1\}.
\end{equation}

\begin{figure}[htbp]
\centering
\includegraphics[width=1.0\textwidth]{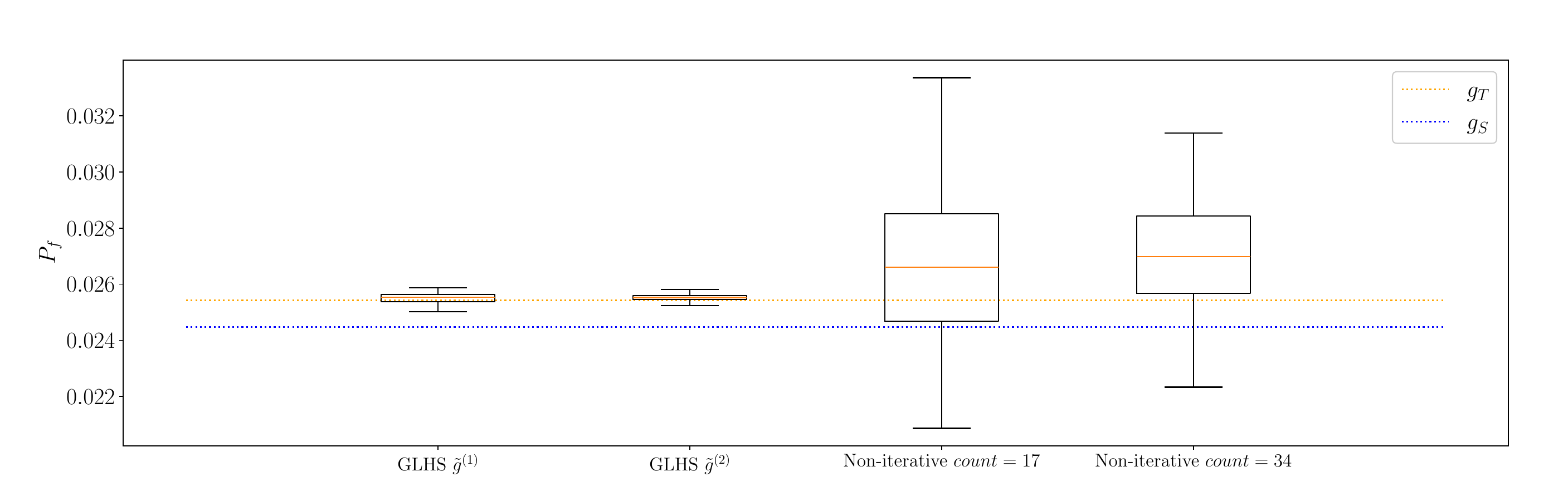}
\caption{Comparison of GLHS with the non-iterative method. Values are given in Table \ref{Table2D_Li}.}
\label{fig:box_plot}
\end{figure}

The corresponding results are shown in Figure~\ref{fig:box_plot} and Table~\ref{Table2D_Li}. $10^7$ Monte Carlo samples are generated, and the same sample set is used to evaluate both $g_S$ and $g_T$. The GLHS and non-iterative methods are each executed 100 times to compute the reported mean and standard deviation.

The results demonstrate that the proposed method achieves accurate failure probability estimates using a limited number of simulations when compared to the non-iterative approach. While no significant improvement is observed in the mean estimate $\mu_{P_f}$ when an additional 17 evaluations are introduced, a consistent reduction in the standard deviation $\sigma_{P_f}$ is observed for both methods, indicating improved estimator stability.

\begin{table}[htbp]
\centering
\begin{tabular}{l c c c c } 
 \hline
  Method & Plot & $\mu_{P_f}$ & $\sigma_{P_f}$ & $m_{T}$ \\ 
 \hline
 Monte Carlo \ref{subsec:MC} & $g_T$ & 0.0254 & - & - \\
 Surrogate \ref{subsec:RS} & $g_S$ & 0.0245 & - & 40 \\
 GLHS \ref{section:LGM} & GLHS $\tilde{g}^{(1)}$ & 0.0255 & $2.03e-04$ & $40+17$ \\
 GLHS \ref{section:LGM} & GLHS $\tilde{g}^{(2)}$ & 0.0255 & $1.76e-04$ & $40+17+17$ \\
 Non-iterative \ref{subsec:NIM} & Non-iterative $\text{count} = 17$ & 0.0268 & $2.80e-03$ & $40+17$ \\ 
 Non-iterative \ref{subsec:NIM} & Non-iterative $\text{count} = 34$ & 0.0271 & $ 1.96e-03$ & $40+17+17$ \\  
 \hline
\end{tabular}
\caption{Results for comparison between the different methods are presented here. It's worth noting that for this comparison, a Monte Carlo simulation was conducted using 10 million samples, with seeding for the computation of $g_T$ and $g_S$. The buffer zone of $g_S$ comprises 23,133 samples. The GLHS was sampled $10^7$ times. The value $m_{\text{hf}}$ represents the number of samples computed for each result. For both methods 40 samples were needed to compute the global surrogate. The mean and standard deviation were obtained by running GLHS and the non-iterative method 100 times.} 
\label{Table2D_Li}
\end{table}

\subsection{Chemical reaction behind a shock}

PLATO (PLAsmas in Thermodynamic Non-Equilibrium) is a simulation tool for modeling multi-component plasmas and is used here to study chemical reactions occurring behind a shock during atmospheric entry into Titan’s atmosphere \citep{doi:10.2514/6.2023-3490}. In this configuration, simulations are conducted in one spatial dimension, with chemical species concentrations reported as functions of the distance behind the shock. In the present work, PLATO serves as a computationally inexpensive model for assessing the performance of the GLHS framework in a higher-dimensional input setting. 
To assess convergence of the failure probability estimate, a total of $10^7$ simulations are performed, leveraging the relatively low computational cost of PLATO. For clarity of visualization, figures in this subsection are generated using a reduced subset of samples, while all tabulated results are computed using the full set of $10^7$ simulations.

The uncertain inputs considered in this study correspond to the pre-exponential factors of three chemical reactions—reactions 60, 2, and 1—selected based on prior sensitivity analyses \citep{doi:10.2514/6.2004-2455, doi:10.2514/6.2022-3576}. In addition, uncertainty in the free-stream methane concentration $X_{\mathrm{CH}_4}$ is included. The chemical reactions considered in this analysis are summarized in Table~\ref{tab:selected_reactions}.
\begin{table}[htbp]
\centering
\caption{Selected dissociation and exchange reactions from \cite{gokcen2007}.}
\label{tab:selected_reactions}
\begin{tabular}{llll}
\toprule
 \textbf{Reaction} & \textbf{$A$ (cc/mol/s)} & \textbf{UF} & $T_a (\text{K})$ \\
\midrule

N$_2$ + M $\rightarrow$ N + N + M
& $7.00 \times 10^{21}$ & 3 & $113,200$\\

N$_2$ + M $\rightarrow$ N + N + M
& $3.00 \times 10^{22}$ & 5 & $113,200$\\

C + N$_2$ $\rightarrow$ CN + N
& $5.24 \times 10^{13}$ & 2 & $4,045$\\
\bottomrule
\end{tabular}
\end{table}

The pre-exponential factor $A_i$ associated with each reaction $i$ is initially characterized by a confidence interval of the form
\begin{equation}
    A_i \in \left[ \frac{A_{\mathrm{ref},i}}{F_i},\, A_{\mathrm{ref},i} F_i \right],
\end{equation}
corresponding to a $99.6\%$ confidence level. Following \cite{Doostan_2013}, $A_i$ is modeled as a lognormally distributed random variable,
\begin{equation}
    A_i = A_{\mathrm{ref},i}\exp\!\left(
    \frac{\log(F_i)}{\Phi^{-1}\!\left(1-\frac{\theta}{2}\right)}\, z_i
    \right),
\end{equation}
where $z_i \sim \mathcal{N}(0,1)$. Here, $A_{\mathrm{ref},i}$ denotes the nominal reference value of the $i$th reaction, and $F_i$ is the associated uncertainty factor. The pre-exponential factor enters the forward reaction rate model through
\begin{equation}
    k_{f,i} = A_i\, T_c^{\,n} \exp\!\left(-\frac{T_a}{T_c}\right),
\end{equation}
where $A_i$ is the pre-exponential factor, $n$ is the temperature exponent, and $T_a$ is the activation temperature. In the two-temperature nonequilibrium formulation, the control temperature $T_c$ depends on the reaction type. For dissociation reactions, $T_c = \sqrt{T\,T_{ve}}$, where $T$ is the translational temperature and $T_{ve}$ denotes the vibrational--electronic temperature. For exchange reactions, the control temperature reduces to $T_c = T$.

Although the uncertainty in $A_i$ is naturally modeled using a lognormal distribution, the current GLHS implementation supports only uniformly distributed inputs. Consequently, the pre-exponential factors are mapped to equivalent uniform distributions over the same confidence intervals. This treatment preserves the original uncertainty bounds and is consistent with previous studies, which also model uncertainty in the free-stream methane concentration $X_{\mathrm{CH}_4}$ using a uniform distribution.

Accordingly, the analysis is performed using a four-dimensional vector of independent, uniformly distributed random variables $\boldsymbol{\xi} \in [-1,1]^4$. These normalized inputs are sampled to construct the PCE, while the corresponding physical parameters used in the PLATO simulations are obtained via the transformation
\begin{equation}
    \begin{aligned}
        A_i &= \exp(\tilde{\mu}_i + \tilde{\sigma}_i \xi_i), \\
        \tilde{\mu}_i &= \frac{\ln(A_{\max,i}) + \ln(A_{\min,i})}{2}, \\
        \tilde{\sigma}_i &= \frac{\ln(A_{\max,i}) - \ln(A_{\min,i})}{2},
    \end{aligned}
\end{equation}
with $A_{\max,i} = A_{\mathrm{ref},i} F_i$ and $A_{\min,i} = A_{\mathrm{ref},i}/F_i$.

PLATO simulations are performed using the pre-exponential factors defined above. From each simulation, three quantities of interest (QoIs) are extracted:
\begin{itemize}
    \item the distance $y$ behind the shock,
    \item the CN species concentration $X_{\mathrm{CN}}$,
    \item the temperature $T_{\mathrm{ve}}$, corresponding to slowly relaxing energy modes, primarily vibrational modes, and including contributions from free-electron translational energy and excitation of bound electrons in heavy particles.
\end{itemize}

\subsubsection{Methodology of running PLATO simulations and post-processing}

PLATO simulations are executed in batches of $N_{\mathrm{batch}} = 5\times10^{5}$ realizations. To ensure statistical independence, the random seed used for uniform sampling over $[-1,1]^4$ is varied between batches. Each batch is distributed across 20 processors, with each processor executing 25{,}000 simulations. For each simulation, the spatial coordinate $y$ behind the shock, the CN concentration $X_{\mathrm{CN}}$, and the temperature $T_{\mathrm{ve}}$ are extracted to compute the QoI $f$ defined as
\begin{equation}
    f(y,\bm{x}) = X_{\mathrm{CN}}(y,\bm{x})\, T_{\mathrm{ve}}^{4}(y,\bm{x}).
\end{equation}
Here, the right-hand side is introduced as a radiation-relevant proxy rather than a direct radiative heat-flux model. 
Specifically, $X_{\mathrm{CN}}(y,\bm{x})$ acts as a measure of the abundance of the dominant radiator (CN), while $T_{ve}(y,\bm{x})$ represents the nonequilibrium vibrational--electronic temperature that controls excitation and emission. 
The product $X_{\mathrm{CN}}\,T_{ve}^4$ therefore provides a scalar indicator of the relative magnitude of CN-driven radiative emission/heating along the post-shock relaxation layer, which is known to be critical for Titan/Dragonfly entry.

In this application, the output associated with a given input realization $\bm{x}_i$ consists of discrete samples of $f$ along the spatial coordinate $y$. To obtain a scalar ground-truth quantity suitable for reliability analysis, we define
\begin{equation}
    \check{g}_T(\bm{x}_i) = \max_{y}\, f(y,\bm{x}_i).
\end{equation}
Rather than directly selecting the maximum among discrete values over $y$, a cubic spline is fitted to $f(y)$, and the maximum value of the spline is used to compute $\check{g}_T$. This procedure mitigates discretization effects and provides a smoother and more robust estimate of the peak response. 

\subsubsection{Results} 

The numerical results presented in this subsection are obtained using the algorithm parameters summarized in Table~\ref{Table4DPlatoinputs}. In this test case, the probability of failure is prescribed as $P_f = 1\%$. To define the corresponding failure threshold, the scalar outputs $\check{g}_T$ evaluated over the full sample set are sorted in descending order, and the upper $1\%$ quantile is identified as the failure region. This procedure defines a threshold value $\check{g}_{\mathrm{lim}}$ separating safe and failed realizations.

\begin{table}[h]
\centering
\begin{tabular}{l l c} 
 \hline
  Parameter & Description & Value \\ 
 \hline
 $m_{c}$ & Number of samples & $10000000$ \\ 
 $c_{lim}$ & Data failed threshold & $0.99$ \\ 
 $d$ & Number of dimensions & 4 \\ 
 $c$ & Constant controlling the buffer zone volume & 1 \\
 $\alpha$ & \% of data used to create initial buffer zone & 0.3 \\
 $m_{0}$ & Number of initial samples to compute $g_S$ & 70 \\ 
 $n$ & Order of global surrogate & 2 \\
 $n_{max}$ & Maximum order of local surrogate & 3 \\
 $m_{l}$ & Number of samples to compute $g_L^{(1)}$ & 56 \\
 $c_{r}$ & Constant to proportionally add samples in the buffer zone & 1.5 \\
 \hline
\end{tabular}
\caption{Initial parameters set by the user for the 4D Chemical reaction test case.}
\label{Table4DPlatoinputs}
\end{table}

The data are subsequently post-processed so that exceedance of the prescribed threshold corresponds to negative values, yielding a consistent limit-state formulation defined as
\begin{equation}
    g_T(\bm{x}) = \check{g}_{\mathrm{lim}} - \check{g}_T(\bm{x}),
\end{equation}
where, for a prescribed level $c_{\mathrm{lim}} \in (0,1)$, the threshold 
$\check{g}_{\mathrm{lim}}$ is defined as the empirical $c_{\mathrm{lim}}$-quantile 
of Monte Carlo realizations of the scalar response $\check{g}_T$. 
In this work, $c_{\mathrm{lim}} = 0.99$.
This construction ensures that realizations with $g_T(\bm{x}) \le 0$ are classified as failures, consistent with the convention adopted throughout the algorithm. By prescribing a target failure probability of $P_f = 1\%$, the failure domain is expected to contain approximately $10^5$ samples out of the $10^7$ available realizations. Since the full input space is accessible in this test case, it is possible to analytically estimate the minimum number of misclassified samples implied by the discrepancy the true and estimated failure probability. This provides a diagnostic criterion for assessing whether the chosen value of $m_d$ is sufficiently conservative, as it must at least account for this expected number of misclassifications.

The global surrogate $g_S$ is constructed using 70 randomly selected samples. A relatively low polynomial order is sufficient, as an accurate fit is achieved at an early stage. The resulting failure probability estimates are reported in Table~\ref{Table4DPf}. Using the global surrogate alone yields a relative error of approximately $3\%$ with respect to the failure probability. 

Consistent with the trends observed in the one- and two-dimensional test cases, the four-dimensional PLATO example exhibits clear convergence of the surrogate-based estimate toward the ground-truth probability of failure. In this test case, only a single GLHS iteration is required: while the global surrogate alone yields a relative error of approximately $3\%$, the combined global–local surrogate reduces this error to $0.01\%$. These results demonstrate that GLHS can achieve rapid convergence and high accuracy in higher-dimensional settings with minimal additional model evaluations.

\begin{table}[h]
\centering
\begin{tabular}{l c c c} 
 \hline
  Model & $\mu_{P_f}$ & $\%$ error with $g_T$ & $m_{\text{hf}}$ \\ 
 \hline
 \(g_{T}\) & 0.010000 & - & - \\ 
 \(g_{S}\) & 0.010300 & $3.00\%$ & 70  \\ 
 \({\tilde{g}}^{(1)}\) & 0.010001 & $0.01\%$ & $70+56$\\  
 \hline
\end{tabular}
\caption{Results for the chemical reaction test case.  
Here, the buffer zone holds 67,390 samples, and the order selected for the local by cross-validation is 2. The value of $\mu_{P_f}$ was obtained by running 10 simulations, yielding a standard deviation of $8.1e-05$ for $\tilde{g}^{(1)}$.}
\label{Table4DPf}
\end{table}

\begin{figure}[htbp]
\centering
\includegraphics[width=1.0\textwidth]{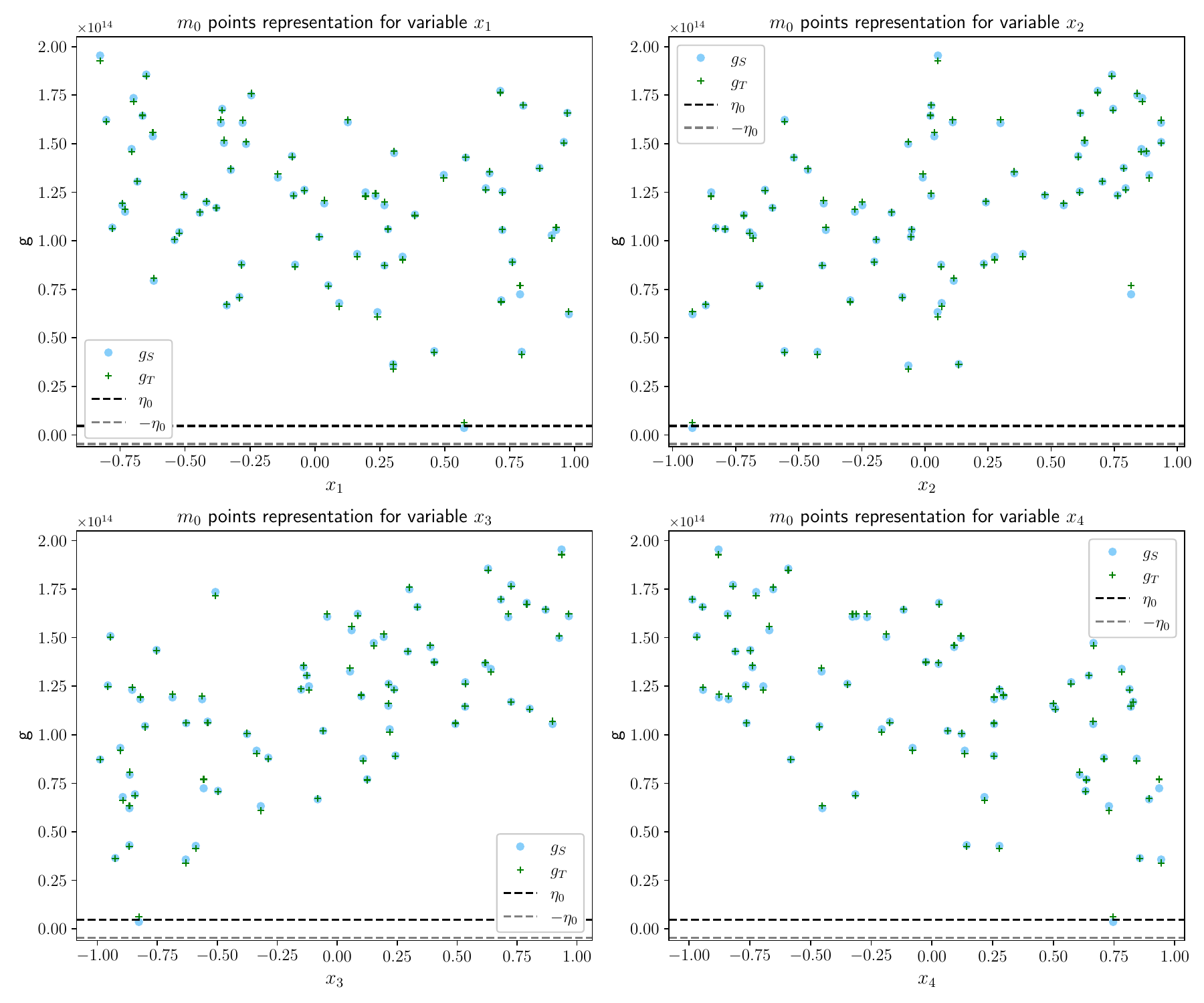}
\caption{The $m_0$ samples are represented with the ground truth and the global surrogate (trained on those samples) for each variable. The buffer zone area is represented by the horizontal line $\pm \eta_0$. Negative values are failed. Note that the current 0 is a presentation for a failure probability at $P_f=0.001$ and $c=1.0$. Only one point can be observed in the buffer zone, and no samples are available in the failure area when the global surrogate was trained.}
\label{fig:m0points}
\end{figure}

\begin{figure}[htbp]
\centering
\includegraphics[width=1.0\textwidth]{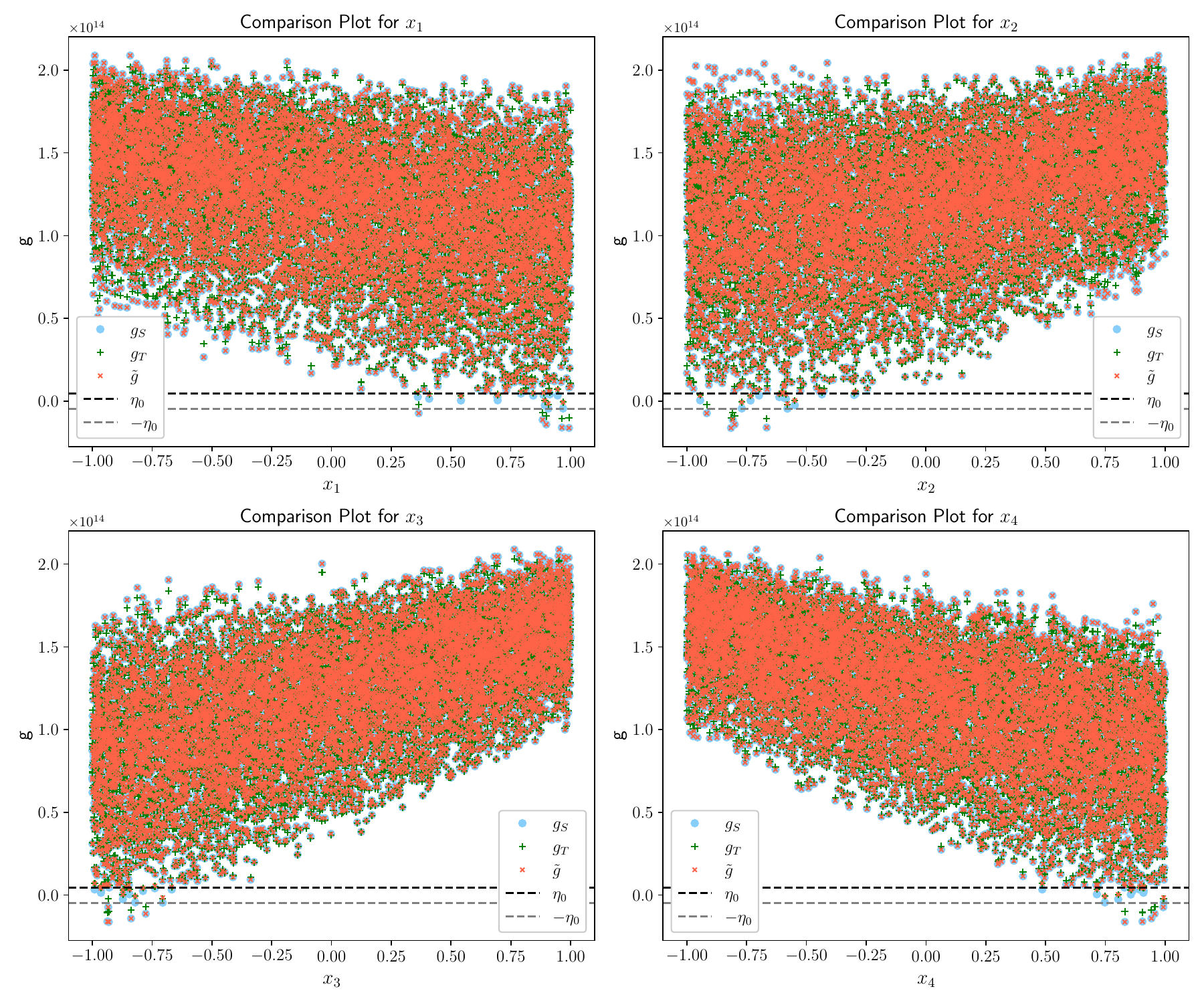}
\caption{A representation of $10000$ samples out of the $10$ million simulations that are used for reference by Monte Carlo estimation. The buffer zone area is represented by the horizontal line $\pm \eta_0$. Negative values are failed. Note that the current 0 is a presentation for a failure probability at $P_f=0.001$ and $c=1.0$.}
\label{fig:m1points}
\end{figure}

\begin{figure}[htb]
\centering
\includegraphics[width=1.0\textwidth]{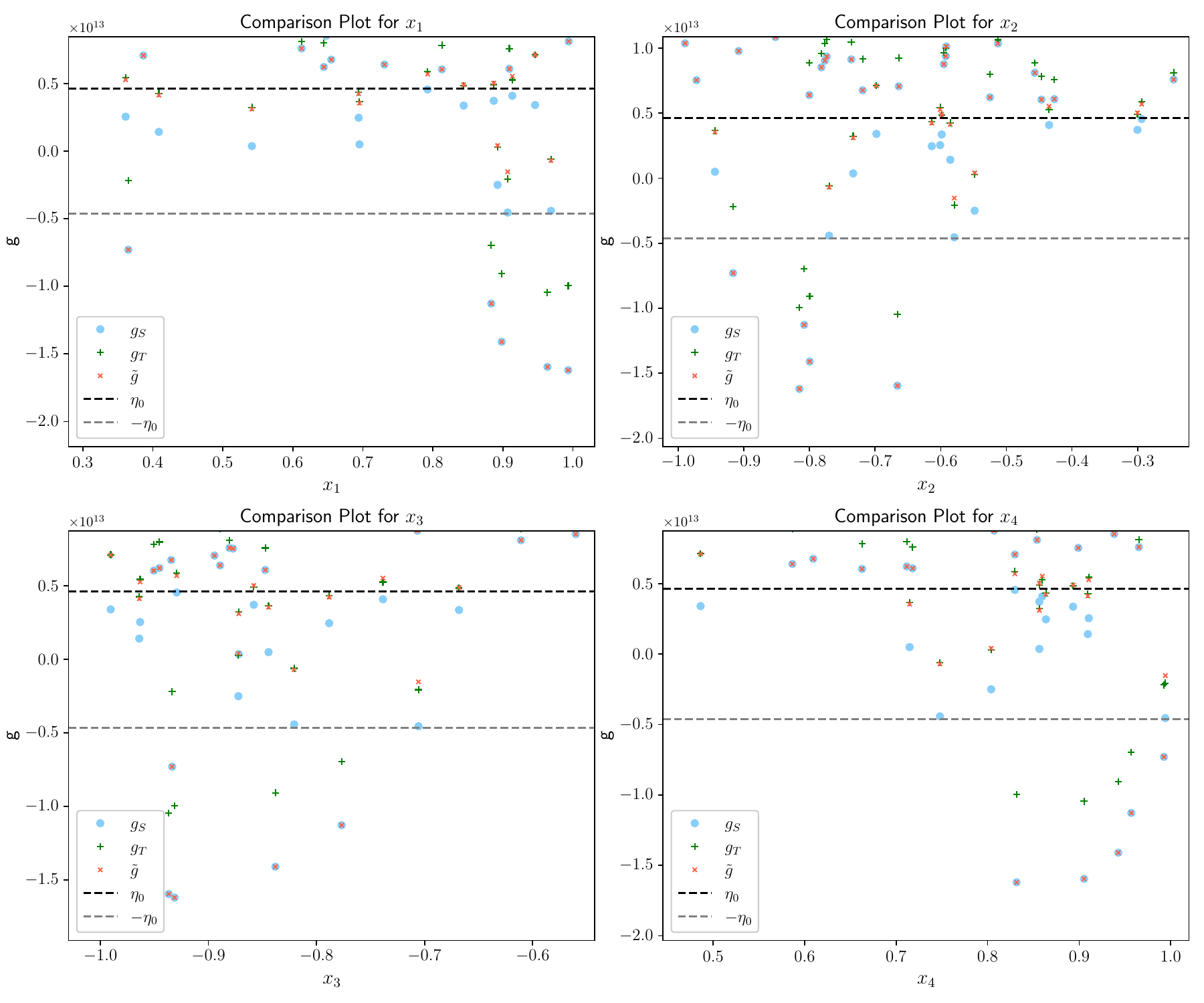}
\caption{A zoom on the buffer zone of Figure \ref{fig:m1points}. The buffer zone area is represented by the horizontal line $\pm \eta_0$. Negative values are failed. Note that the current 0 is a presentation for a failure probability at $P_f=0.001$}
\label{fig:m1pointszoom}
\end{figure}

To further investigate the behavior of the surrogate models, results are also reported for a smaller failure probability $P_f = 0.1\%$. The samples used to construct the global surrogate are shown in Figure~\ref{fig:m0points}. While the global surrogate $g_S$ provides an overall accurate approximation of $g_T$, localized discrepancies are observable in regions associated with failure. Figure~\ref{fig:m1points} compares the ground-truth function $g_T$, the global surrogate $g_S$, and the combined surrogate $\tilde{g}^{(1)}$, using $10{,}000$ samples drawn from the full set of $10^7$ simulations. A zoomed view of the buffer zone is provided in Figure~\ref{fig:m1pointszoom}. These figures show that the local surrogate significantly improves the approximation within the buffer zone, where accurate resolution is critical for estimating low failure probabilities. All failure probability estimates, however, are computed using the full set of $10^7$ samples.

The corresponding quantitative results are summarized in Table~\ref{Table4DPfb}, using the same algorithmic setup as in Table~\ref{Table4DPlatoinputs}. While the global surrogate yields a relative error of $3.0\%$ for $P_f = 1\%$, this error increases substantially to $49.5\%$ when estimating $P_f = 0.1\%$. This behavior reflects a known limitation of global surrogate models in rare-event settings, where failure realizations may appear as outliers or remain insufficiently resolved. In contrast, the introduction of a local surrogate reduces the relative error from $49.5\%$ to $0.9\%$, demonstrating the effectiveness of the GLHS framework in accurately capturing low-probability failure events.

\begin{table}[htbp]
\centering
\begin{tabular}{l c c c} 
 \hline
 Model & $\mu_{P_f}$ & $\%$ error with $g_T$ & $m_{\text{hf}}$ \\ 
 \hline
 \(g_{T}\) & 0.001000 & - & - \\ 
 \(g_{S}\) & 0.001495 & $49.5\%$ & 70  \\ 
 \({\tilde{g}^{(1)}}\) & 0.001002 & $0.2\%$ & $70+56$\\  
 \hline
\end{tabular}
\caption{Results for the chemical reaction test case. 
The probability of failure is a mean value from 10 simulations, with a standard deviation of $4.9e-06$ for $\tilde{g}^{(1)}$. Note that the buffer zone holds 16,647 samples, and the order selected for the local through cross-validation is 2.}
\label{Table4DPfb}
\end{table}

Based on these results, a sensitivity study is performed to assess the effect of the buffer zone conservativeness parameter $c$. When $c$ is decreased, corresponding to a less conservative buffer zone, an increasing number of misclassified samples may be excluded, which can lead to a biased failure probability estimate. Conversely, increasing $c$ yields a more conservative buffer zone that fully encloses the transition region of the ground-truth limit-state function. A larger initial buffer zone also enables the possibility of performing additional GLHS iterations if required. The impact of varying $c$ for a single iteration of the algorithm is summarized in Table~\ref{Table4Dc}. While variations in $c$ do not result in a significant change in the mean failure probability estimate $\mu_{P_f}$, an increase in the standard deviation is observed as the buffer zone expands. This behavior is expected, as a larger buffer zone requires the local surrogate to approximate the response over a broader region. If the polynomial order of the surrogate is not increased accordingly, the approximation quality may degrade. In addition, the number of samples $m_d$ within the buffer zone increases with $c$, since the analysis is performed on a fixed grid of $10^7$ samples, and enlarging the buffer zone naturally increases the number of samples selected.

\begin{table}[htbp]
\centering
\begin{tabular}{l c c c c c} 
 \hline
  & $\mu_{P_f}$ & $\%$ error with $g_T$ & $\sigma_{P_f}$ & c & $m_d$\\ 
 \hline
 \(g_{T}\) & 0.001000 & - & - & - & -\\ 
  \({\tilde{g}}^{(1)}\) & 0.001002 & $0.2\%$ & 4.9e-06 & 1.0 & 16647\\ 
 \({\tilde{g}}^{(1)}\) & 0.001003 & $0.3\%$ & 7.6e-06 & 2.0 & 35016\\  
  \({\tilde{g}}^{(1)}\) & 0.000999 & $0.1\%$ & 1.2e-05 & 3.0 & 56441\\ 
 \hline
\end{tabular}
\caption{Failure probability estimation for the chemical reaction test case for different $c$ values. 
}
\label{Table4Dc}
\end{table}

\section{Conclusion}
\label{sec:con}

This work presents the development and application of the Global-Local Hybrid Surrogate (GLHS) method for efficient and accurate estimation of failure probabilities in high-dimensional systems. The method integrates a global surrogate, constructed from a limited set of high-fidelity simulations, with iteratively refined local surrogates focused around the estimated failure domain. By employing Christoffel Adaptive Sampling (CS) within dynamically learned buffer zones, GLHS effectively targets regions of uncertainty and refines surrogate accuracy without high computational cost.

The numerical results across one-, two-, and four-dimensional test cases demonstrate that GLHS
consistently improves the estimation of failure probability compared to standard surrogate
methods and the non-iterative sampling strategy of~\cite{LI20108966}. In the 1D test case, the use of
a local surrogate reduced the failure probability error from 6.8\% to 0\%, while in the 2D case,
the error decreased from 3.5\% to 0.4\% using only 17 additional simulations. The
method also maintains strong performance in more challenging settings, including a 4D
reliability analysis involving PLATO an entry simulations.

Beyond accuracy, GLHS offers notable computational advantages by concentrating model
evaluations near the transition regions surrounding the limit state surface, thereby avoiding the
cost of full-domain resampling. Nevertheless, several aspects of the current framework may be
further improved. The present implementation assumes a single transition region (or buffer
zone), but more complex systems may feature multiple disconnected failure domains. A possible
extension is to incorporate clustering-based domain learning to automatically identify and
refine multiple buffer zones. Another promising improvement is to employ optimal sampling
measures tailored to the surrogate construction; for instance, when using an
\(\ell_1\)-type reconstruction, such as LASSO, sampling according to optimal measures could further
enhance stability and efficiency.

Additionally, we aim to extend the methodology to accommodate Gaussian or other non-uniform input distributions by adopting generalized polynomial bases (e.g., Hermite polynomials) and adjusting the CS measure accordingly. Such extensions would broaden the applicability of GLHS to real-world systems characterized by complex uncertainty distributions.

\section*{Acknowledgments}
This work has been supported under a NASA Space Technology Research Institute Award (ACCESS, grant number 80NSSC21K1117). 
\newpage
\appendix

\section{Nomenclature}
\label{appendix:nomenclature}

\begin{longtable}{||p{4.2cm}|p{7.0cm}|p{3.5cm}||}
\caption{Comprehensive notation table for the GLHS methodology.} \\
\hline
\textbf{Category} & \textbf{Description} & \textbf{Variable} \\
\hline
\endfirsthead

\hline
\textbf{Category} & \textbf{Description} & \textbf{Variable} \\
\hline
\endhead

\hline
\multicolumn{3}{r}{\emph{Continued on next page}} \\
\endfoot

\hline
\endlastfoot

\multirow{11}{*}{General}
& Input dimensionality & $d$ \\ \cline{2-3}
& Domain bounds & $[\underline{x}, \overline{x}]$ \\ \cline{2-3}
& Input random vector & $\bm{x} = (\bm{x}_1,\dots,\bm{x}_d)$ \\ \cline{2-3}
& Domain of input variables & $\mathcal{D}$ \\ \cline{2-3}
& Prior input distribution & $\tau$ \\ \cline{2-3}
& Restriction of prior measure & $\tau_1$ \\ \cline{2-3}
& Number of samples in the global domain & $m_K$ \\ \cline{2-3}
& Monte Carlo sample count & $m_c$ \\ \cline{2-3}
& Initial samples for global surrogate & $m_0$ \\ \cline{2-3}
& Samples at iteration $l$ & $m_l$ \\ \cline{2-3}
& Norm & $\ell$ \\ \cline{2-3}
& Norm value & $p$ \\ 
\hline

\multirow{8}{*}{Limit-state \& Surrogates}
& Ground-truth limit-state (HF simulation) & $g_T(\bm{x})$ \\ \cline{2-3}
& Global surrogate & $g_S(\bm{x})$ \\ \cline{2-3}
& Local surrogate at iteration $l$ & $g_L^{(l)}(\bm{x})$ \\ \cline{2-3}
& Hybrid global-local surrogate & $\tilde{g}^{(l)}(\bm{x})$ \\ \cline{2-3}
& Failure domain set & $\mathcal{S}_f$ \\ \cline{2-3}
& Buffer zone set & $\mathcal{S}_\eta$ \\ \cline{2-3}
& Probability of failure & $P_f$ \\ \cline{2-3}
& Indicator function & $\mathbb{I}$ \\
\hline

\multirow{8}{*}{Thresholds}
& Initial threshold defining buffer zone & $\eta_0$ \\ \cline{2-3}
& Threshold at iteration $l$ & $\eta_l$ \\ \cline{2-3}
& Conservative expansion constant & $c$ \\ \cline{2-3}
& Fraction of samples used to compute $\eta_0$ & $\alpha$ \\ \cline{2-3}
& Resampling expansion factor & $c_r$ \\ \cline{2-3}
& Hyperrectangle padding & $\varepsilon$ \\ \cline{2-3}
& Set of point selected to compute the surrogate error & $\mathcal{Y}$ \\ \cline{2-3}
& Iteration index & $l$ \\
\hline

\multirow{16}{*}{Sampling \& Sets}
& Global grid & $\mathcal{X}$ \\ \cline{2-3}
& buffer zone grid (iteration $l$) & $\mathcal{X}_l$ \\ \cline{2-3}
& Grid constructed by domain-learning method & $\mathcal{S}_d$ \\ \cline{2-3}
& CS sampling set & $\mathcal{S}_l$ \\ \cline{2-3}
& Generic sample set & $\mathcal{S}$ \\ \cline{2-3}
& Target DL resampling size & $m_d$ \\ \cline{2-3}
& CS QR sample size & $m_f$ \\ \cline{2-3}
& Batch size for hyperrectangle resampling & $m_r$ \\ \cline{2-3}
& Number of CS-selected local HF samples & $m_l$ \\ \cline{2-3}
& Number of samples in the set $\mathcal{X}_l$ & $m_h$ \\ \cline{2-3}
& Total HF samples used & $m_{\text{hf}}$ \\ \cline{2-3}
& Sampling measure based on prior & $\nu$ \\ \cline{2-3}
& Index set of buffer zone samples at iteration $l$ & $\mathcal{I}_l$ \\ \cline{2-3}
& Index set of buffer zone samples & $\mathcal{I}$ \\ \cline{2-3}
& Hyperrectangle & $\mathcal{R}$ \\ \cline{2-3}
& Dirac distribution & $\delta$ \\
\hline

\multirow{9}{*}{Polynomial Approx. (PCE)}
& Global PCE order & $n$ \\ \cline{2-3}
& Maximum local PCE order & $n_{\max}$ \\ \cline{2-3}
& Locally selected optimal PCE order & $n_l$ \\ \cline{2-3}
& Number of PCE basis functions & $N$ \\ \cline{2-3}
& Hyperbolic-cross index set & $\Lambda$ \\ \cline{2-3}
& (Non-orthonormal) Legendre basis & $\psi_j$ \\ \cline{2-3}
& Orthonormal basis after QR & $\phi_j$ \\ \cline{2-3}
& Polynomial approximation spaces & $P_\psi,\; P_\phi$ \\ \cline{2-3}
& Coefficient vector & $\mathbf{c}$ \\
\hline

\multirow{10}{*}{Christoffel Adaptive Sampling}
& Discrete weights in $\tau_l$ & $\rho_i$ \\ \cline{2-3}
& Samples from $\tau_l$ & $\tilde{\bm{x}}_i$ \\ \cline{2-3}
& Measurement matrix & $\mathbf{B}$ \\ \cline{2-3}
& QR factors & $\mathbf{Q},\mathbf{R}$ \\ \cline{2-3}
& Reciprocal Christoffel function & $K_C(P_\phi)(\bm{x})$ \\ \cline{2-3}
& Christoffel sampling weight & $w(\bm{x})$ \\ \cline{2-3}
& Sampling measure for CS & $\nu$ \\ \cline{2-3}
& Number of features & $n_f$ \\ \cline{2-3}
& Basis-size sampling rule & $m_i \sim N_l\log N_l$ \\ \cline{2-3}
& Orthonormal transform coefficients & $q_{kj}$ \\
\hline

\multirow{5}{*}{Domain Learning}
& SVM weight vector for hyperplane & $\mathbf{W}$ \\ \cline{2-3}
& SVM intercept & $b$ \\ \cline{2-3}
& Hyperrectangle bounding region & $\mathcal{R}$ \\ \cline{2-3}
& Upper/lower supporting hyperplanes & $H_U,\ H_L$ \\ \cline{2-3}
& Local hyperplane & $H(\mathbf{z})$ \\ \cline{2-3}
\hline

\multirow{1}{*}{Statistics}
& Mean & $\mu$ \\ \cline{2-3}
& Standard deviation & $\sigma$ \\ \cline{2-3}
& Indicator function & $\mathbb{I}$ \\
\hline

\multirow{7}{*}{PLATO Chemistry}
& Pre-exponential factor (reaction $i$) & $A_i$ \\ \cline{2-3}
& Reference pre-exponential factor & $A_{\mathrm{ref},i}$ \\ \cline{2-3}
& Uncertainty factor & $F_i$ \\ \cline{2-3}
& Vibrational-electronic temperature & $T_{\mathrm{ve}}$ \\ \cline{2-3}
& CN concentration & $X_{\mathrm{CN}}$ \\ \cline{2-3}
& Methane concentration & $X_{\mathrm{CH}_4}$ \\ \cline{2-3}
& Empirical quantile for temperature threshold & $c_\text{lim}$ \\ \cline{2-3}
& Quantity of interest: $X_{\mathrm{CN}} T_{\mathrm{ve}}^4$ & $f(y)$ \\
\hline

\end{longtable}

\section{Algorithm for Domain Learning}
\label{appendix:DL}
To analyze its computational complexity of the domain learning algorithm, we evaluate each stage separately.

\textbf{Stage (a)} involves evaluating the surrogate function at \( m_K \) sample points, leading to a complexity of \( \mathcal{O}(m_K) \).

\textbf{Stage (b)} offers two strategies depending on the dimensionality:
\begin{itemize}
    \item \textit{Hyperplane approach}: Computing a separating hyperplane using a support vector machine (SVM) with a linear kernel has complexity \( \mathcal{O}(n_f \times m_K) \), where \( n_f \) is the number of features and \( m_K \) is the number of points. In our case, \( n_f = 3 \), so the complexity becomes \( \mathcal{O}(3m_K) \).
    \item \textit{Hypercube approach}: Computing the minimum and maximum values across each of the \( d \) dimensions requires \( m_K \times d \) operations, yielding a complexity of \( \mathcal{O}(m_K \times d) \).
\end{itemize}

\textbf{Stage (c)} performs resampling within the selected region and has similar complexity for both the hyperplane and hyperrectangle methods. Generating a single random number in the interval \([0, 1]\) has complexity \( \mathcal{O}(1) \), so generating \( m_r \) samples in \( d \) dimensions yields a total complexity of \( \mathcal{O}(m_r \times d) \).

\textbf{Stage (d)} involves Christoffel Adaptive Sampling, which includes:
\begin{itemize}
    \item constructing the measurement matrix: \( \mathcal{O}(m_r) \),
    \item performing QR factorization: \( \mathcal{O}(m_f \times N) \),
    \item and generating samples from the resulting measure.
\end{itemize}
Here, \( N \) is the number of polynomial basis functions used to construct the measurement matrix (typically \( N \approx \mathcal{O}(10^2) \)). Thus, this stage has an overall complexity of \( \mathcal{O}(m_r + m_f \times N) \).

\medskip

Summing the contributions from all stages, the total complexity is:
\[
\mathcal{O}(m_K + m_l) + \mathcal{O}(m_K d + m_r d + m_f N) \approx \mathcal{O}(m_K d),
\]
assuming \( m_K \gg m_l \), \( m_K \approx \mathcal{O}(10^r) \), \( m_f \approx \mathcal{O}(c \times 10^r) \) with \( c \in [0, 1) \), and \( m_r \approx \mathcal{O}(10^{r-1}) \).

\begin{tcolorbox}[floatplacement=!htbp,float,title=Method: Domain Learning]
\noindent \textbf{Inputs:}  Finite grid $\mathcal{X}=\{\bm{x}_i\}_{i=1}^{m_K}$ representing $\mathcal{D}$, probability measure $\tau$ over $\mathcal{X}$, sampling numbers $m_l,m_r,m_d$, surrogate function $\tilde{g}$ computed on $\bm{x}$, threshold $\eta>0$.
\\
\noindent \textbf{Initialize:} Set $\mathcal{X}_0 = \mathcal{X}$, for $l$-th iteration do\\ 
\\
\textbf{Stage (a):} Compute potential buffer zone $\mathcal{X}_l$ defined as 
\begin{equation*}
    \mathcal{X}_l = \{\bm{x}_i \in \mathcal{X}_0: |\tilde{g}^{(l-1)}(\bm{x}_i)|\leq \eta_{l-1}\}.
\end{equation*}
\textbf{Stage (b):} Compute hyperrectangle/hyperplanes
\begin{itemize}
    \item Hyperplane: For case $d=1$, compute a linear hyperplane $H_l$ via SVM which separates the data $\mathcal{X}_l$ and $\bm{x} \backslash \mathcal{X}_l$. Take the two support vectors $H_U + \varepsilon$ and $H_L - \varepsilon$ and construct the hyperplanes generated by these vectors. Next, compute 
    \begin{equation*}
    a_1 = (H_U + \varepsilon) \cap \mathcal{X}_l,\quad
    b_1 = (H_L - \varepsilon) \cap \mathcal{X}_l.
    \end{equation*}
    Now, the domain $\mathcal{X}_l$ should be contained in the interval $\mathcal{R}= [a_1, b_1]$.
    \item hyperrectangle: For case $d\geq 2$, consider a sample $\bm{x}_i=(x^i_1,\ldots,x^i_d)\in \mathcal{X}_0$, then we compute the minimum and maximum value for each component throughout the grid $\mathcal{X}_0$.
\begin{equation*}
    a_j = \min_{i=1,\ldots,K} x^i_j + \epsilon ,\quad
    b_j = \max_{i=1,\ldots,K} x^i_j + \epsilon ,\quad
    \forall j=1,\ldots,d,
\end{equation*}
where $\epsilon>0$ is a small quantity to expand the limits. Next, construct the hyperrectangle by taking the rectangle created by the limits $a_j,b_j$ as 
\begin{equation*}
    \mathcal{R} = [a_1,b_1] \times [a_2,b_2] \times \ldots \times [a_d,b_d], \quad
    \mathcal{X}_l \subseteq R.
\end{equation*}
\end{itemize}

\textbf{Stage (c):} Resample on hyperrectangle.
\begin{itemize}
    \item[] For each $j=1,\ldots,d$ component, generate $m_r$ uniform random samples $\bm{x}^a_i$ in $[0,1]$, then transform into values in $[a_j,b_j]$. Write $\mathcal{X}^a=\{\bm{x}^a_i\}_{i=1}^{m_r}$, compute the additional grid $\mathcal{X}_l^a$ and merged it to $\mathcal{X}_l$, then select the first $m_d$ samples as  
    \begin{equation*}
    \mathcal{X}_l^a =\{\bm{x}^a_i\in \mathcal{X}^a: |\tilde{g}^{(l-1)}(\bm{x}_i^a)|\leq \eta_{l-1}\},\quad 
    \mathcal{X}_l = \mathcal{X}_l\cup \mathcal{X}_l^a,\quad
    \mathcal{X}_l = \{\bm{x}_i\}_{i=1}^{m_d}.
    \end{equation*} 
\end{itemize}
\textbf{Stage (d):} Draw samples via CS.
\begin{itemize}
    \item[] First, construct sampling measure $\nu$ via CS on $\mathcal{X}_l$.
    \item[] Second, draw $m_l$ samples $\bm{x}_1,\ldots,\bm{x}_{m_l}$ via CS, in other words, $\bm{x}_i\sim\nu$ for $i=1,\ldots,m_l$.
\end{itemize} 
\textbf{Outputs:} Potential buffer zone $\mathcal{X}_l=\{\bm{x}_i\}_{i=1}^{m_h}$ and samples $\bm{x}_{1},\ldots,\bm{x}_{m_l}$ on $\mathcal{X}_l$.
\end{tcolorbox}

\begin{tcolorbox}[floatplacement=!htbp,float,title= Method: GLHS]
\noindent \textbf{Inputs:} Constant values required by the algorithm are set. This includes values such as $c$, $\alpha$, $m_{0}$, $m_{c}$, the maximum order of the local surrogate, the order of the global surrogate, etc.\\

\noindent \textbf{Initialize:} Sample strategy method to compute $\bm{x} \in \mathbb{R}^{d \times m_{d}}$, $\bm{x}_{C} \in \mathbb{R}^{d \times m_{c}}$, $\eta_0$, $\bm{x}_{0}$, $g_T(\bm{x}_{0})$, the global surrogate $g_S$ and the reference values for $P_f$ based on $g_T$ and $g_S$ (if available).\\

\textbf{while $\mathcal{X}_l\neq \emptyset$ do} \\
\begin{itemize}
    \item[] \textbf{Stage (a):} Domain learning method is called to compute the hyperectangle on the buffer zone $\mathcal{X}_l$ based on $\eta_{l-1}$, generate samples, then use CS to extract $m_l$ points. Refer to appendix \ref{appendix:DL} concerning that section details.\\

    \item[] \textbf{Stage (b):} Truth $g_T$ is computed for the $m_l$ points.\\  

    \item[] \textbf{Stage (c):} Local surrogate $g_L^{(l)}$ is computed for optimum order.\\ 

    \item[] \textbf{Stage (d):}  Compute new $\eta_{l}$ to create the new buffer zone. Set $l = l + 1$ and continue.\\
\end{itemize}

\textbf{Outputs:} The combination of global and local surrogate $\tilde{g}^{(l)}(\bm{x})$. For the local surrogate, the index of the samples used, the $R$ matrix from the QR decomposition, the matrix of the coefficients $\bm{c}$, the $\bm{\eta}$ values that give the buffer, and the optimum order. The probability of failure can also be computed by the algorithm as output. 
\end{tcolorbox}  

\clearpage
\bibliographystyle{elsarticle-harv} 
\bibliography{example}






\end{document}